\documentclass{aa}  

\usepackage{graphicx}
\usepackage{siunitx}
\usepackage{booktabs}
\usepackage{bm}
\usepackage{txfonts}
\usepackage{hyperref}
\hypersetup{colorlinks=true, linkcolor=red, urlcolor=black, citecolor=blue}
\usepackage{array,multirow}
\usepackage{float}
\newcommand*\diff{\mathop{}\!\mathrm{d}}
\usepackage{gensymb}

\usepackage[dvipsnames]{xcolor}

\usepackage{natbib}
\bibpunct{(}{)}{;}{a}{}{,} 

\begin{document}

   \title{
   Reconstructing the extended structure of multiple sources strongly lensed by the ultra-massive elliptical galaxy SDSS~J0100+1818}

\author{A. Bolamperti\inst{1,2} \thanks{\email{andrea.bolamperti@phd.unipd.it}}
    \and C. Grillo\inst{3,4}
    \and R. Ca\~nameras\inst{5}
    \and S. H. Suyu\inst{5,6,7}
    \and L. Christensen\inst{8,9}
    }

\institute{Dipartimento di Fisica e Astronomia, Università degli Studi di Padova, Vicolo dell'Osservatorio 3, I-35122 Padova, Italy 
     \and INAF -- Osservatorio Astronomico di Padova, Vicolo dell'Osservatorio 5, I-35122, Padova, Italy \label{inafpd}
     \and Dipartimento di Fisica, Università degli Studi di Milano, via Celoria 16, I-20133 Milano, Italy
     \and INAF-IASF Milano, via A. Corti 12, I-20133 Milano, Italy
     \and Max-Planck-Institut f{\"u}r Astrophysik, Karl-Schwarzschild-Str.~1, 85748 Garching, Germany \label{mpa} \goodbreak
     \and  Technical University of Munich, TUM School of Natural Sciences, Department of Physics, James-Franck-Str. 1, 85748 Garching, Germany \label{tum} \goodbreak
     \and Academia Sinica Institute of Astronomy and Astrophysics (ASIAA), 11F of ASMAB, No.1, Section 4, Roosevelt Road, Taipei~10617, Taiwan \label{asiaa} \goodbreak
     \and Cosmic Dawn Center (DAWN), Denmark \goodbreak
    \and Niels Bohr Institute, University of Copenhagen, Jagtvej 128, 2200 Copenhagen N, Denmark \goodbreak
}

   \date{Received --; accepted --}

 
  \abstract
  {We study the total and baryonic mass distributions of the deflector SDSS\,J0100+1818 through a full strong lensing analysis. The system is composed by an ultra-massive early-type galaxy at $z=0.581$, with total stellar mass of $(1.5 \pm 0.3) \times 10^{12}$~M$_\odot$ and stellar velocity dispersion of ($450 \pm 40$) km s$^{-1}$, surrounded by ten multiple images of three background sources, two of which spectroscopically confirmed at $z=1.880$. We take advantage of high-resolution \textit{HST} photometry and VLT/X-shooter spectroscopy to measure the positions of the multiple images and perform a strong lensing study with the software {\tt GLEE}. We test different total mass profiles for the lens and model the background sources first as point-like and then as extended objects. We successfully predict the positions of the observed multiple images and reconstruct over approximately 7200 \textit{HST} pixels the complex surface brightness distributions of the sources. 
  We measure the cumulative total mass profile of the lens and find a total mass value 
  of $(9.1 \pm 0.1) \times 10^{12}$~M$_\odot$, within the Einstein radius of approximately 42~kpc, and stellar-over-total mass fractions ranging from ($49 \pm 12$)\%, at the half-light radius ($R_e = 9.3$~kpc) of the lens galaxy, to ($10 \pm 2$)\%, in the outer regions ($R = 70$~kpc). These results suggest that the baryonic mass component of SDSS\,J0100+1818 is very concentrated in its core and that the lens early-type galaxy/group is immersed in a massive dark matter halo, which allows it to act as a powerful gravitational lens, creating multiple images with exceptional angular separations. This is consistent with what found in other ultra-high mass candidates at intermediate redshift.
  We measure also the physical sizes of the distant sources, resolving them down to a few hundreds of parsec. Finally, we quantify and discuss a relevant source of systematic uncertainties on the reconstructed sizes of background galaxies, associated to the adopted lens total mass model.
  }


   \maketitle
\section{Introduction}

In the last decades, several studies have suggested the presence of a significant amount of dark matter (DM), both in elliptical \citep[e.g.,~][]{Loewenstein2002} and spiral \citep[e.g.,~][]{Rubin1983} galaxies. These studies have measured total mass-to-light ratios considerably higher than the stellar ones and observed a remarkable difference between the visible, i.e. stellar plus gas, and total mass estimated from galaxy dynamics \citep[e.g.,][]{Rubin1970, Gerhard2001, Cappellari2006}.
Since then, cosmological observations of the cosmic microwave background \citep[CMB;][]{Smoot1992, Hinshaw2013, PlanckCollaboration2020}, of baryon acoustic oscillations \citep[BAO;][]{Efstathiou2002, Eisenstein2005, DESCollaboration2021} and of Type Ia supernovae \citep[SNe;][]{Riess1998, Perlmutter1999, Scolnic2018} have all supported the currently accepted $\Lambda$CDM model, according to which the Universe is composed of baryons and cold dark matter for $ \approx \! 30\%$ and of a poorly understood component, known as dark energy and responsible of the accelerated expansion of the Universe, for the remaining $ \approx \! 70\%$. 
However, the $\Lambda$CDM model, which has been very successful at large ($\gtrsim \! 1$ Mpc) scales, cannot predict accurately the properties of structures at smaller scales, like the value of the inner slope of dark matter halos \citep[e.g.,][]{Gnedin2004, Newman2013a, Newman2013b, Martizzi2012}. Hence, the interplay between dark matter and baryons on galactic scales is still being intensively investigated. 

$N$-body simulations of DM particles show that, at equilibrium, they are distributed following an almost universal mass density profile, $\rho(r)$, first described by the Navarro-Frenk-White (NFW; \citealt*{Navarro1997}) profile. This profile has a characteristic slope, which varies from $\rho_{r\ll r_s} \propto r^{-1}$, at small radii, thus with a central cusp, to $\rho_{r\gg r_s} \propto r^{-3}$, at large radii. Recent simulations with higher resolution \citep[e.g.,][]{Golse2002, Graham2006, Navarro2010, Gao2012, Collett2017, Dekel2017} and with realistic models describing the baryonic components predict mass density profiles which deviate from the NFW one. This mainly depends on processes such as gas cooling, which allows baryons to condense towards the center of a galaxy \citep[e.g.,][]{Blumenthal1986, Gnedin2004, Sellwood2005, Gustafsson2006, Pedrosa2009, Abadi2010, Sommer-Larsen2010}, active galactic nuclei (AGNs) feedback \citep[e.g.,][]{Peirani2008, Martizzi2013, Li2017}, dynamical heating in the central cuspy region, due to infalling satellites and mergers \citep[e.g.,][]{El-Zant2001, El-Zant2004, Nipoti2004, Romano-Diaz2008, Tonini2006, Laporte2015}, and thermal and mechanical feedback from supernovae \citep[e.g.,][]{Navarro1996b, Governato2010, Pontzen2012}. \\

In this context, the observation of very massive (and dark-matter rich) galaxies and the measurement of their inner total, dark matter and baryonic mass profiles represent a key step in the comprehension on how the different mass components are distributed, and to infer how galaxies formed and evolved over cosmic time. These studies can be addressed by exploiting very massive galaxies which act as gravitational lenses.

Strong gravitational lensing represents a powerful tool for studying galaxy evolution and for exploring the properties of the Universe \citep[e.g.~][]{Bartelmann2010, Treu2010}. Given the fact that the deflection of the light emitted by a source only depends on the total gravitational potential of the lens, gravitational lensing is sensitive to both luminous and DM mass, and allows one to obtain some of the most precise total mass measurements, with relative errors of the order of a few percent, in extragalactic astrophysics \citep[e.g.~][]{Grillo2010, Zitrin2012}. 
Together with the results obtained by exploiting the most massive lenses, i.e. galaxy clusters, about the slope value of the mass density of the dark matter halos in their inner regions \citep{Sand2004, Grillo2015, Annunziatella2017, Bergamini2019}, the measurements of the cosmological density parameter values \citep{Jullo2010, Caminha2016, Grillo2020}, and the value of the Hubble constant \citep{Grillo2018}, galaxy-scale strong gravitational lenses have provided several key results (e.g., \citealt{Suyu2013, Suyu2017}). 
In particular, it has been possible to characterize the physical properties of the lens, for instance to reconstruct the total and dark-matter mass distributions, thus the dark-matter over total mass fraction \citep{Gavazzi2007, Grillo2009, Suyu2010, Sonnenfeld2015, Schuldt2019} and the mass density slopes \citep[e.g.,][]{Treu2002, Koopmans2009, Barnabe2011, Shu2015}, to infer the most likely lens stellar initial mass function \citep[IMF; e.g.,][]{Canameras2017a, Barnabe2013, Sonnenfeld2019}, and to identify dark-matter substructures \citep[e.g.,][]{Vegetti2012, Hezaveh2016, Ritondale2019}. 
It has also been shown how to measure the values of some cosmological parameters in strong lensing systems with kinematic data of the lenses \citep[e.g.,][]{Grillo2008, Cao2012} or in systems where two or more sources are multiply imaged by the same lens galaxy \citep{Tu2009, Collett2014, Tanaka2016, Smith2021}, once the mass sheet degeneracy is broken \citep{Schneider2014}. 

Moreover, gravitational lensing offers the opportunity to study the lensed background sources: they can be fully reconstructed considering their surface brightness distribution during the strong lensing modeling \citep[e.g.,][]{Schuldt2019, Rizzo2021, Wang2022}, or analyzed in detail thanks to the high magnification factors, which allow one to investigate the local feedback mechanisms driving the evolution of the faintest and smallest high-$z$ galaxies. \citep[e.g.][]{ForsterSchreiber2009, Canameras2017b, Cava2018, Iani2021}

In this paper, we measure the total and baryonic mass distributions of the deflector SDSS\,J010049.18+181827.7 (hereafter, SDSS\,J0100+1818), included in the Cambridge And Sloan Survey Of Wide ARcs in the skY (CASSOWARY) survey \citep{Belokurov2009, Stark2013}\footnote{The ID of this system has changed in the CASSOWARY survey in the past years. It was called CSWA\,15 when we first targeted it, but it is currently identified as CSWA\,115. In the paper, to avoid confusion, we will refer to it as SDSS\,J0100+1818.}. The exceptionally large Einstein radius of $\approx \! 42 \, \si{kpc}$, together with the results from our follow-up observations with the Nordic Optical Telescope (NOT), the Very Large Telescope (VLT) and the Hubble Space Telescope ({\it HST}), presented in Section \ref{sec:data}, suggest that the SDSS\,J0100+1818 deflector is an uncommon strong lens and a candidate fossil system at intermediate redshift \citep{Johnson2018}. 
%

This paper is organized as follows. In Sect.~\ref{sec:data}, we present the photometric and spectroscopic observations of SDSS\,J0100+1818. In Sect.~\ref{sec:system}, we focus on the main lens galaxy, showing the results on its stellar mass, kinematics, luminosity profile and environment. In Sect.~\ref{sec:point}, we perform a strong lensing analysis by exploiting different sets of point-like multiple images, and illustrate the best-fit models and the deflector total mass profile. We enhance the analysis in Sect.~\ref{sec:extended}, where we model the deflector by considering the extended surface brightness of the lensed sources. At the end of this section, we study the reconstructed sources assuming different models and comparing their reconstructed sizes. In Sect.~\ref{sec:discussion}, we summarize and discuss the main results. 
Throughout this work, we assume $H_0 = 70 \, \si{km.s^{-1}.Mpc^{-1}}$, $\Omega_m = 0.3$ and $\Omega_\Lambda = 0.7$. In this model, $1$ arcsec corresponds to a linear size of $6.59$ kpc at the deflector redshift of $z = 0.581$. 

\section{Observations and data reduction}
\label{sec:data}

\begin{figure}
   \centering
   \includegraphics[width=8cm]{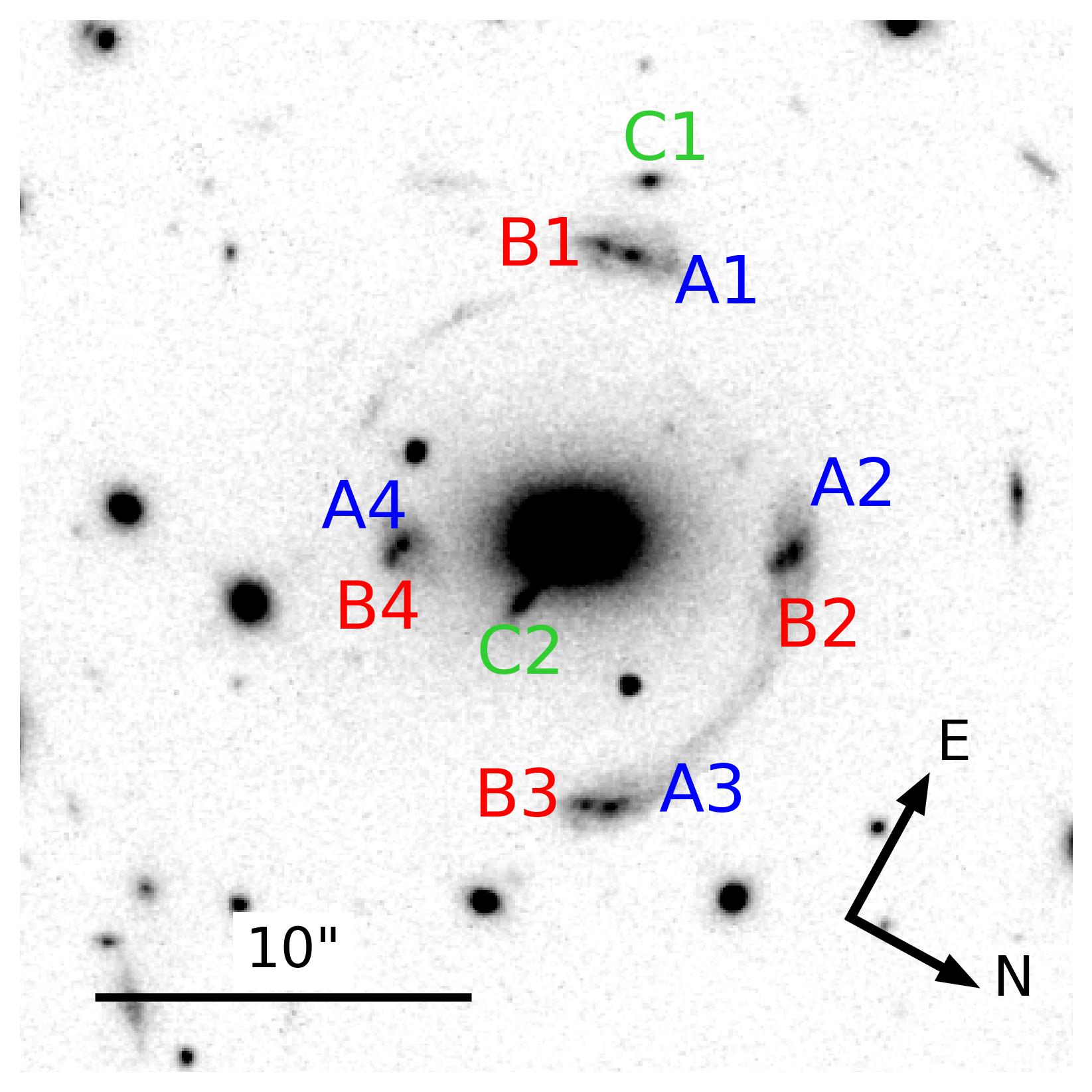}
      \caption{F160W image of the strong lensing system analyzed in this work. We labeled with A1-A4, B1-B4, and C1-C2 the multiple images corresponding to each of the three background sources considered in the analysis. Later in this study, we will show that sources A and B represent two emission peaks of a single extended source.          }
         \label{fig:labels}
   \end{figure}

\subsection{HST imaging}
We observed SDSS\,J0100+1818 with the {\it HST} Wide Field Camera 3 (WFC3) in June and August 2018 (program GO-15253; PI: Ca\~nameras), during one orbit in each of the two F438W and F160W filters. We used {\tt astrodrizzle} from the {\tt DrizzlePac} software package \citep{Fruchter2010}  to improve the correction for geometric distortion on the individual exposures calibrated with the standard {\it HST} pipeline. We also optimized both the rejection of cosmic rays with dedicated bad pixel masks, and the subtraction of the local sky background. Small regions with decreased sensitivity in the IR channel of WFC3 (blobs) were corrected using flat fields from the most recent version of the IR blob monitoring \citep{Sunnquist2018}. The individual exposures were then redrizzled and combined with inverse-variance weighting. For the F160W band, we adopted a pixel scale 0.066\arcsec\ pix$^{-1}$ to adequately sample the PSF, and a value  ${\rm {\tt final\_pixfrac} = 0.8}$ optimized for this pixel scale. For the F438W band, we used 0.033\arcsec\ pix$^{-1}$ and ${\rm {\tt final\_pixfrac} = 0.6}$. The PSF FWHMs measured on isolated stars in the field are 0.086\arcsec\ and 0.187\arcsec\ in F438W and F160W, respectively.

\subsection{NOT imaging}
We used the 2.5\,m Nordic Optical Telescope (NOT; program 56-032, PI: Ca\~nameras), to obtain color information for galaxies in the lens environment and to characterize their spectral energy distributions. In October 2017, we obtained StanCam imaging in the B and R bands with good or average weather conditions, and total exposure times of 2400 and 3600~s, respectively. We added a 900~s long V-band exposure obtained with ALFOSC in October 2013 through the fast track service (program 47-426, PI: Grillo). Image reduction was conducted with standard IRAF routines. We then used the Scamp and SWarp softwares \citep{Bertin2006,Bertin2010} to correct for geometric distorsion, resample individual frames and align them with respect to the WCS, using the USNO-B1 catalog as reference, and to obtain relative alignments with a rms accuracy below 0.1\arcsec. This resulted in a joint field-of-view of 2.5\arcmin\,$\times$\,2.5\arcmin, pixel sizes of 0.19\arcsec, and PSF FWHMs of 1.05\arcsec, 0.70\arcsec, and 0.95\arcsec, in B, V, and R, respectively. Photometric calibration was performed with respect to SDSS, using color corrections from \citet{Jester2005}, and refined by fitting blackbody functions to stars in the field. The AB magnitudes of the main lens galaxy in B, V, and R bands are 22.57$\pm$0.06, 21.62$\pm$0.10, and 20.22$\pm$0.12, respectively.

\begin{figure}
  \centering
  \includegraphics[width=0.50\textwidth]{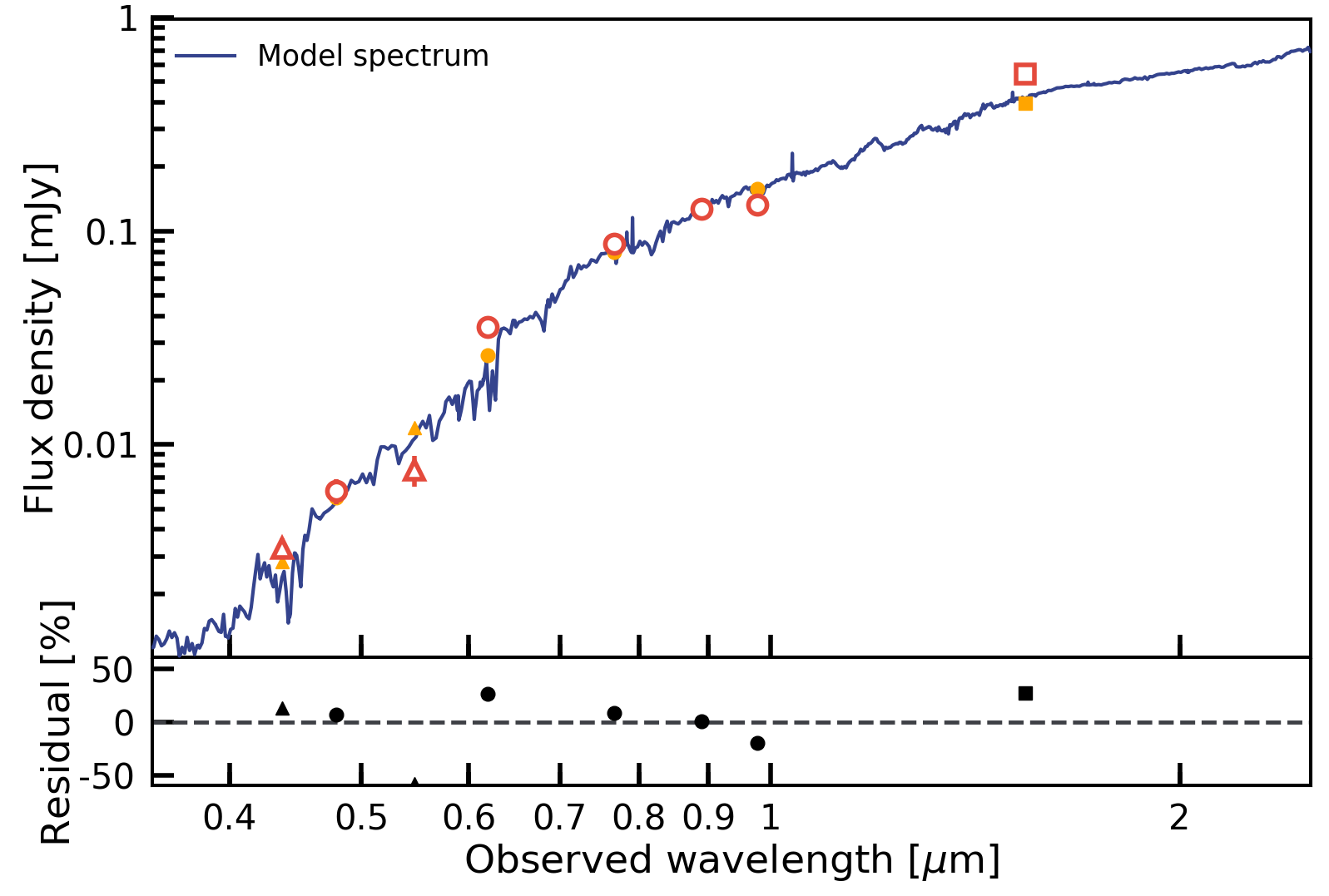}
  \caption{Spectral energy distribution of the main lens galaxy of SDSS\,J0100+1818. The observed flux densities from NOT (red triangles), PanSTARRS (red circles) and {\it HST} (red square) are used to fit the SED with the CIGALE software. Errorbars are smaller than the symbols. The best-fit stellar continuum template (blue curve) corresponds to the model fluxes plotted in orange, and is used to infer the total stellar mass value of the deflector. The bottom panel shows the relative residuals of the fit.}
  \label{fig:cigale}
\end{figure}

\subsection{VLT/X-Shooter spectroscopy}
SDSS\,J0100+1818 was observed with VLT/X-Shooter \citep{Vernet2011} (program 091.A-0852, PI: Christensen), in order to measure the lens and source spectroscopic redshifts, and to infer the lens stellar velocity dispersion. The main lens galaxy was observed in August and September 2013 with a 1.2\arcsec\ wide slit, seeing $\simeq$1\arcsec, clear sky conditions, and with a total on-source integration time of 3.2~hours aimed at reaching a S/N of 5$-$10 per spectral bin in the VIS arm. The five individual OBs used a generic object, sky, object observing sequence with a 21\arcsec\ offset. As part of the same program, a separate one-hour long OB targeted two multiple images of the brightest lensed source, by nodding between the two images with a 19\arcsec\ offset.

After rejecting cosmic rays with {\tt Astro-SCRAPPY} \citep{Mccully2019}, the OBs were reduced with the X-Shooter pipeline \citep{Modigliani2010} using the EsoReflex environment version 2.8.3 \citep{Freudling2013}. The OB nodding between the two lensed images was reduced in stare mode in order to measure the sky background directly from the frames. We used the observations of telluric standard stars taken within two hours from the science observations at similar airmass, to correct for telluric absorption using the atmospheric synthetic spectra from {\tt Molecfit} \citep{Smette2015,Kausch2015}.
The combination of individual exposures, the subtraction of residual sky emission, and the optimization of the wavelength calibration were then performed with separate scripts \citep[see][]{Selsing2019}. The 0.1$-$1\arcsec\ offsets between the UVB and VIS arms resulting from the lack of atmospheric dispersion compensation at the time of the observations were also corrected on the 2D spectra. Finally, we conducted an optimal, profile-weighted extraction of the 1D spectrum of the lens galaxy of SDSS\,J0100+1818, and we extracted 1D spectra of multiple images using fixed apertures.

\section{The SDSS\,J0100+1818 deflector}
\label{sec:system}

SDSS\,J0100+1818, (RA, dec$)=($01:00:49.18,~+18:18:27.79), was introduced in a late version of the CASSOWARY catalog, and was not included in the spectroscopic confirmation program from \citet{Stark2013}. We therefore used the X-Shooter spectra to measure secure redshifts for the main lens elliptical galaxy and lensed source. The best-fit lens redshift of $z=0.581$ was inferred from the most prominent rest-frame optical absorption lines, as part of the stellar kinematic analysis detailed in Sect.~\ref{ssec:vdisp}. We inferred a joint redshift of $z=1.880$ for the two source components forming image families A and B visible in Fig.~\ref{fig:labels}. Given the low S/N values of the lines detected in the 2D spectra, this source redshift estimate is obtained from a joint analysis of the multiple images targeted by the X-Shooter OBs. The width of the lines detected at about $10740 \, \AA$ in the binned 2D spectra of A1/B1, A3/B3, and A4/B4 is consistent with the [OII]$\lambda$$\lambda$3727 doublet. Together with a faint detection of [OIII]$\lambda$5007 in A3/B3 (Appendix \ref{app_a}) and with the lack of additional line detections over the spectral range covered by X-Shooter, the detection of [OII]$\lambda$$\lambda$3727 for the separate images ensures that the source redshift is robust. The 1D spectra of these multiple images are shown in the Appendix \ref{app_a}. 

By combining observations in the \textit{HST} F160W and F438W filters with the first strong lensing models of the system, we identified another candidate background source, whose two multiple images are labeled with C1 and C2 in Fig.~\ref{fig:labels}. Even though we are lacking spectroscopic confirmation, we include the source C in our strong lensing models, with its redshift value as a free parameter, because the observation of two images with similar colors, robustly predicted at the same positions by the strong lensing models makes its multiply-imaged nature highly likely. However, without a spectroscopic redshift measurement for source C, we cannot determine which between AB and C is the most distant source, so we will not be able to perform a multi-plane lensing analysis here, i.e., the light emitted by each source will be deflected only by the total mass distribution of the main lens, and not by any other background source. Then, thanks to the available photometric data and lensing model predictions, we also hypothesize the presence of an additional background source, with four multiple images. These images are barely detected in the \textit{HST} data and two of the multiple images appear angularly very close to candidate group members.
Thus, the one-component mass models we will develop in this study might not be able to properly reconstruct these images. For these reasons, we do not consider this extra background source in the strong lensing models presented in this work, where we will focus on the multiple images A, B, and C. We postpone to a future study a more detailed analysis of this fourth multiple-image family, when new data will possibly become available (see the last point in the summary). 

In the following, we will present the physical properties of the main deflector, i.e., its stellar mass, luminosity profile and stellar kinematics, measured from the full photometric and spectroscopic data set. The last subsection is dedicated to the study of the lens environment, discussing the possible group nature of this system.

\subsection{Stellar mass} 
The spectral energy distribution (SED) of the main lens galaxy is modeled with the Code Investigating GAlaxy Emission \citep[CIGALE,][]{Burgarella2005,Noll2009,Boquien2019}. We use the multiband photometry from PanSTARRS, NOT, and {\it HST}, taking the F160W flux corrected from contamination by neighboring sources. The magnitudes are measured with SExtractor \citep{Bertin1996}, using ISOCORR down to 3$\sigma$ isophotes in all bands for the main lens galaxy, except for the magnitude in {\it HST} F160W filter, measured from its luminosity profile, as described in the next subsection. 
The grid of models to conduct SED fitting relies on \citet{Bruzual2003} single stellar population templates, and assume delayed star formation histories with exponential bursts and e-folding times in the range 0.1--1~Gyr, ages between 0.5 and 8~Gyr, and the modified \citet{Charlot2000} extinction law. We assume a Salpeter \citep{Salpeter1955} stellar initial mass function and fix the metallicity to solar values, as expected for massive early-type galaxies at $z \sim 0.6$ \citep{Sonnenfeld2015,Conroy2013,Gallazzi2014}. The best-fit SED shown in Fig.~\ref{fig:cigale} results in a total stellar mass value of $(1.5 \pm 0.3) \times 10^{12}$~M$_{\odot}$.

\subsection{Luminosity profile} 
\begin{figure*}
   \centering
   \includegraphics[width=0.95\textwidth]{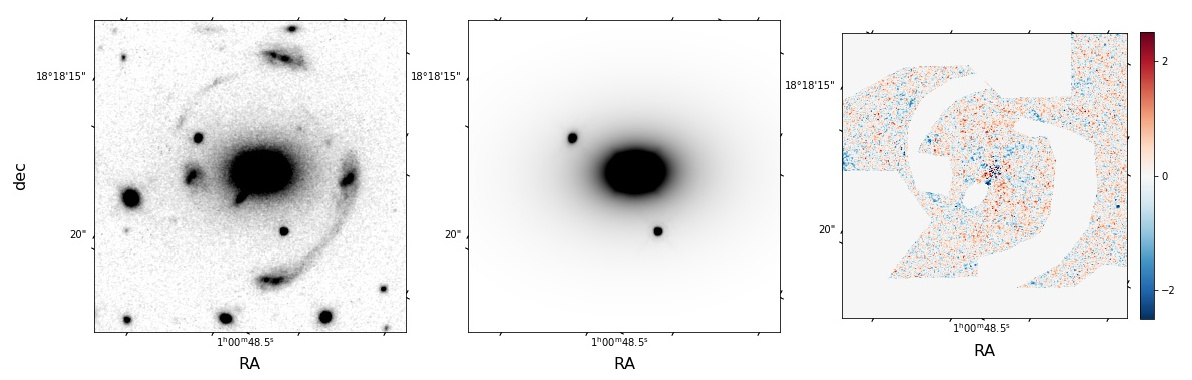}
   \caption{On the left, the considered cutout of the \textit{HST} image in the F160W band. In the center, the best-fit model, consisting of a combination of two Sérsic profiles for the main lens in the middle, and a single Sérsic profile for each of the neighbouring galaxies. On the right, the normalized residuals on a color scale approximately from $-2\sigma$ to $+2\sigma$. The white pixels, with values of $0.0$, show the regions masked during the luminosity profile modeling.}
    \label{photo residuals}%
\end{figure*}

We perform the photometric modeling of the system using the public software GALFIT \citep{Peng2002, Peng2010}. Considering the \textit{HST} image in the F160W band, we extracted a cutout centered on the deflector with size of approximately $30\arcsec \times 30\arcsec$ (see Fig.~\ref{photo residuals}). Then, we decided to model the main lens and two other bright galaxies, located between the main lens and the multiple images/arcs. We masked out the external regions, where some bright objects are present, and the arcs, including the multiple image C2 (see the masked pixels, in white, in the right-hand panel of Fig.~\ref{photo residuals}). In particular, C2 is too close to the main lens and radially distorted to be separately and adequately modeled. The very low values of the normalized residuals, without any clear pattern, confirm the goodness of our modeling choices.

We assume Sérsic \citep{Sersic1963} profiles for the galaxies, and a uniform distribution for the background. The parameter values are optimized by minimizing a standard $\chi^2$ function, defined as
\begin{equation}
    \chi^2 = \sum_i^{N_{\rm pix}} \frac{\left( f_{i}^\mathrm{data}-f_{i}^\mathrm{model} \right) ^2}{\sigma^2_i} \, ,
\end{equation}
where the index $i$ runs over all the non-masked pixels, up to $N_{\rm pix}$, $f_{i}^\mathrm{data}$ is the intensity value of the $i$-th pixel in the input data image and $f_{i}^\mathrm{model}$ is the intensity value in the same pixel predicted by the model, after the convolution with the estimated point spread function (PSF). The PSF we use was obtained by combining several non-saturated bright stars in the field of view, stacking them, subtracting the background, and then normalizing the resulting image. 

To better reproduce the observations, and thus to reduce the residuals in the inner regions, the main lens galaxy has been modeled with a combination of two Sérsic profiles. The centers of the two components are not forced to coincide, but the best-fit centers differ by less than 1 pixel. In Fig. \ref{photo residuals}, the considered cutout is shown in the left panel, the best-fit model in the middle one, and the normalized residuals in the right one. The number of degrees of freedom is evaluated from the number of considered (non-masked) pixels $N_{\rm pix}$ and the number of free parameters of the model $N_{\rm par}$, as ndof~$=N_{\rm pix}-N_{\rm par}$.
The minimum $\chi^2$ value is 13899 
 which, with a number of degrees of freedom ndof~$=46752$, corresponds to a reduced $\chi^2$ value of $\sim 0.3$. 

\begin{figure}
   \centering
   \includegraphics[width=0.48\textwidth]{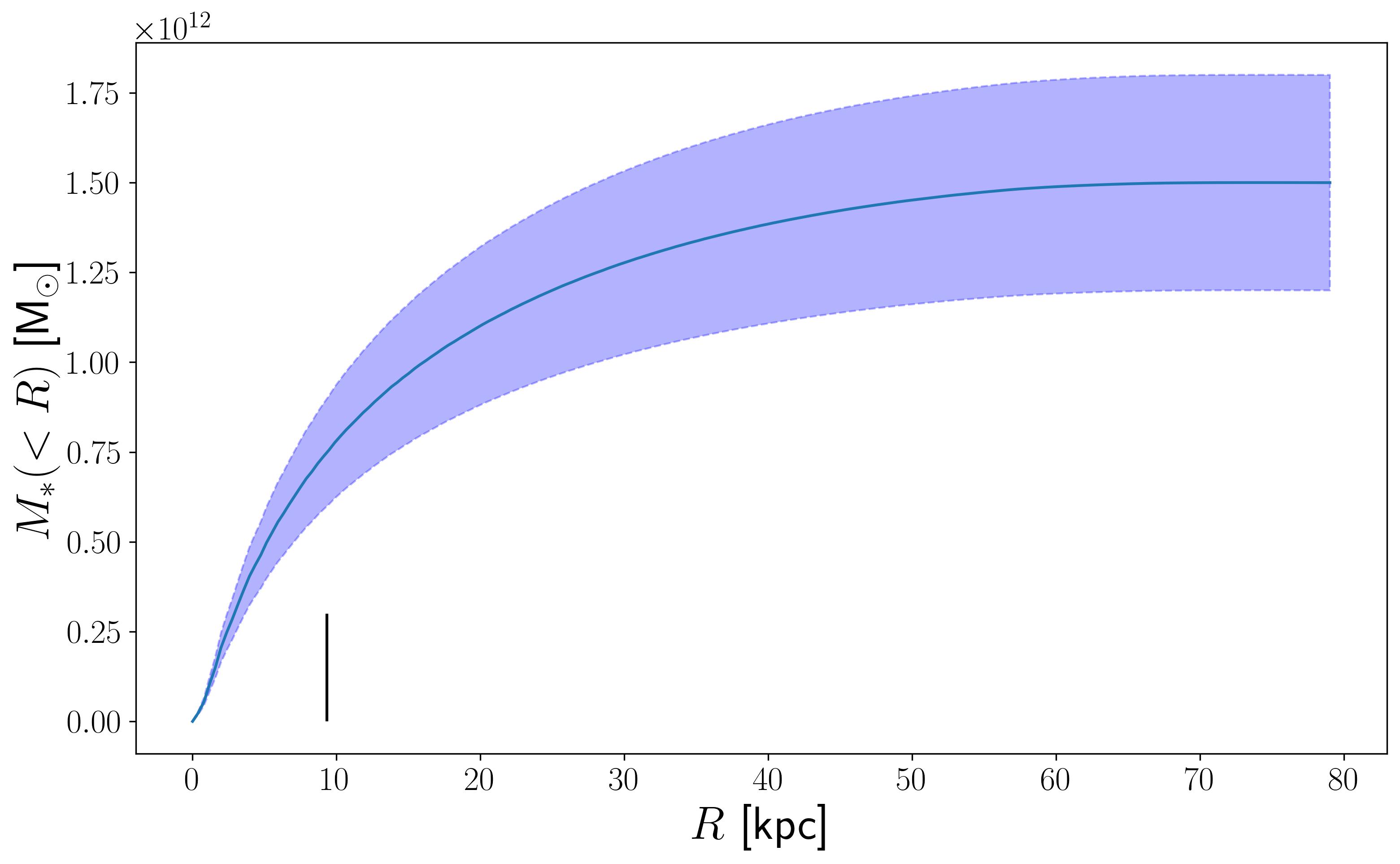}
      \caption{Cumulative stellar mass ($M_*$) profile (blue solid line) and $\pm 1\sigma$ uncertainties (shaded area) of the main lens. It is measured from the total stellar mass measured through the SED fitting and assuming a constant stellar mass-to-light ratio. The black vertical line shows the half-light radius of the lens galaxy.}
    \label{fig:stellarmass_profile}
\end{figure}

We make use of the best-fit model of the main lens galaxy for three purposes: \\
1) we work with lens-subtracted images in the strong gravitational lensing modeling, explained in the following sections, to avoid that the lens light contaminates the multiple images and the arcs. This will be particularly important when considering the source C, for which the multiple image C2 is angularly very close to the lens center; \\
2) we build the cumulative luminosity profile of the deflector, from which we measure a total F160W magnitude value of $m_{F160W} = 17.06 \pm 0.05$, and an effective (i.e., half-light) radius value of $(1.42 \pm 0.02)\arcsec$, corresponding to $(9.32 \pm 0.12)$ kpc at $z=0.581$. 
The value of the total magnitude is considered in the SED fitting (see Fig. \ref{fig:cigale}), from which we measure the total stellar mass value; \\
3) we convert the luminosity profile into a stellar mass profile, showed with a blue solid line in Fig.~\ref{fig:stellarmass_profile}, by assuming a constant stellar mass-to-light-ratio. In detail, we consider the galaxy cumulative luminosity profile, normalize it with the total luminosity value, and then multiply this adimensional profile by the value of the total stellar mass of the central galaxy, obtained from its SED fitting. The shaded area indicates the $1\sigma$ uncertainty, and the black vertical line locates, on the $x$-axis, the value of the effective radius.

\subsection{Stellar kinematics}
\label{ssec:vdisp}
We use the penalized Pixel Fitting software \citep[pPXF,][]{Cappellari2004,Cappellari2017} that follows a maximum penalized likelihood method to determine the stellar velocity dispersion $\sigma_{*}$ of the foreground lens galaxy.  The X-Shooter spectrum has a resolution of $\simeq$6500 in the VIS arm, where the main absorption lines fall. We masked the strong telluric absorption windows that could affect the analysis, and re-binned by two spectral pixels in order to obtain a ${\rm SNR > 10}$ per bin and ensure reliable measurements of the stellar kinematics. We used the SYNTHE high-resolution template library \citep{Munari2005} that covers the entire 2500--10500~\AA\ wavelength range, with 1~\AA~pix$^{-1}$ sampling and a constant resolution FWHM of 2~\AA~pix$^{-1}$. Given the large size of the library, we extracted a subset of spectra representing the overall range of effective temperatures. The template spectra were then matched to the resolution of the binned X-Shooter spectrum. The pPXF fit was conducted in the rest-frame wavelength range 3800-5300~\AA\ that covers the main absorption lines, Ca~H and K at 3935 and 3970~\AA, the G-band at 4305~\AA, H$\beta$ at 4863~\AA, the MgI triplet at 5167, 5173 and 5184~\AA, as well as the strong continuum break at 4000~\AA. This results in a stellar velocity dispersion value of (451 ± 37) km~s$^{-1}$, confirming that the main lens galaxy of SDSS\,J0100+1818 is among the rarest, most massive elliptical galaxies \citep{Loeb2003}.

\subsection{Environment} 
\label{ssec:env}
\begin{figure}
   \centering
   \includegraphics[width=0.48\textwidth]{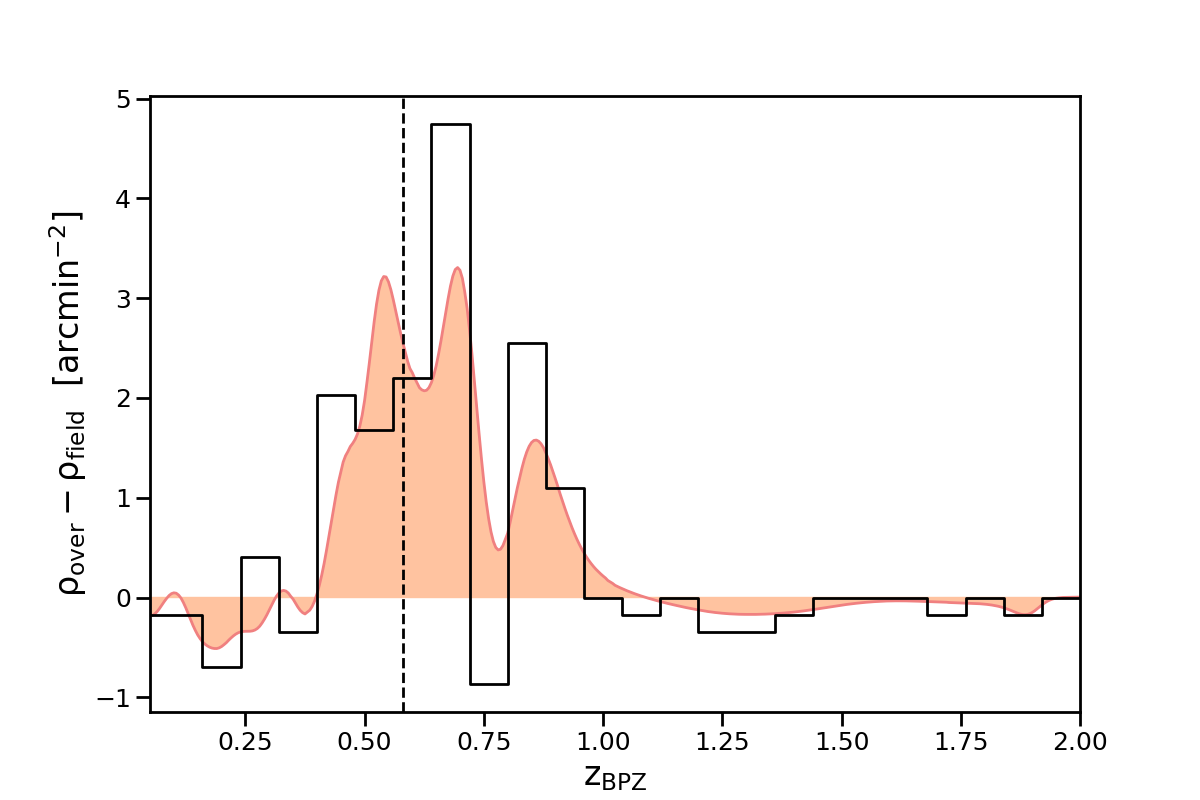}
      \caption{Difference between the redshift distribution functions of galaxies located within 0.5\arcmin\ from the main ultra-massive lens galaxy at $z=0.581$, and over the rest of the 2.5\arcmin\,$\times$\,2.5\arcmin\ field, after normalizing to the same area (orange curve). The black histogram is obtained from the peak of the redshift distribution of each source. A total of 14 galaxies within 0.5\arcmin\ from the main deflector have redshifts consistent, at the 2$\sigma$ confidence level, with $z=0.581$, and 39 over the rest of our analyzed field.}
    \label{fig:zdistrib}
\end{figure}

Given the large angular separation between the multiple images, we must consider the possibility that the main lens galaxy is located within a rich and overdense environment, as for other high-mass lens early-type galaxies at intermediate redshift \citep[e.g.,][]{Newman2015,Johnson2018,Wang2022}. We estimate photometric redshifts using the template-fitting {\tt BPZ} package \citep{Benitez2000} to study the source distribution over the 2.5\arcmin\,$\times$\,2.5\arcmin\, field-of-view covered by our multiband images. In particular, we use the \textit{grizy} bands from PanSTARRS and BVR from NOT. At this stage, to avoid the possible introduction of systematics, we do not include the F160W band from \textit{HST}, which only covers the central part of the considered field. The combination of optical and near-infrared bands cover the 4000\,\AA\ break for early-type galaxies up to $z \sim 1$. We obtain redshift estimates for a total of 142 galaxies over the field, with a typical uncertainty of $\pm 0.2$, and 53 of the most reliable redshifts are consistent at the 2$\sigma$ level with $z=0.581$. Figure~\ref{fig:zdistrib} shows the difference between the redshift probability distribution functions of sources within 0.5\arcmin\ from the main lens (about 200~kpc at $z=0.581$), and over the rest of the field, both normalized to the same area. Our analysis is broadly consistent with the presence of an overdensity around the main deflector, and possible galaxy members mostly have R-band AB magnitudes of 22$-$24~mag. Definite confirmation of this structure nonetheless requires spectroscopic follow-up and, given the large spread of candidate members over the field, we simply model its effect on the light deflection with an external shear component.

\section{Point-like source modeling}
\label{sec:point}
We used the Gravitational Lens Efficient Explorer \citep[{\tt GLEE};][]{Suyu2010, Suyu2012} software to perform our strong lensing modeling of the system. {\tt GLEE} supports several types of mass and light profiles and uses Bayesian analyses, such as simulated annealing and Markov Chain Monte Carlo (MCMC), to infer the probability distributions of the parameter values and thus characterize the deflector. It also employs the Emcee package, developed by \cite{Foreman-Mackey2013}, for sampling the model parameter posterior. 

We start our modeling making use of point-like images, i.e., considering each multiple image position as that of its brightest pixel, with an uncertainty of one pixel.
The best-fit values of the parameters of each model are estimated through a multi-step simulated annealing technique, first minimizing the $\chi^2$ on the source plane, and then on the deflector plane. Furthermore, the median values and the uncertainties (mainly at the 68\% confidence level) are extracted from MCMC chains of $10^6$ total steps, with acceptance rates between $20\%$ and $30\%$, and rejecting the first $10\%$ burn-in steps. The results showed in Tab.~\ref{tab:mcmc} are extracted from the final chains of a sequence, in which each intermediate chain is used to estimate the covariance matrix of the model parameters and to extract the starting point for the following one.

We model the total mass distribution of the lens by assuming different mass profiles. The goodness of each model is evaluated by comparing the number of degrees of freedom (dof) and the value of the minumum $\chi^2$, computed by comparing the observed, $\bm{\theta}_i^\mathrm{obs}$, and model-predicted, $\bm{\theta}_i^\mathrm{pred}$, multiple image positions with the following statistics:
\begin{equation}
    \chi^2 = \sum_{i=0}^N \frac{|\bm{\theta}_i^\mathrm{obs}-\bm{\theta}_i^\mathrm{pred}|^2}{\sigma^2_i} \, ,
\end{equation}
where $N$ is the total number of multiple images and $\sigma_i$ is the positional uncertainty relative to the $i$-th image. To compare different models, we consider three statistical estimators, often employed in similar strong lensing studies \citep[see, e.g.,][]{Acebron2017, Mahler2018, Caminha2022}:
\begin{enumerate}
    \item the root-mean-square (rms) between the observed and model-predicted image positions, defined as 
    \begin{equation}
        \mathrm{rms} = \sqrt{\frac{1}{N}|\bm{\theta}_i^\mathrm{obs}-\bm{\theta}_i^\mathrm{pred}|^2} \, ;
    \end{equation}

    \item the Bayesian Information Criterion \citep[BIC,][]{Schwarz1978}, given by 
    \begin{equation}
        \mathrm{BIC} = k \ln{n} + \chi^2 \, ,
    \end{equation}
    where $k$ is the number of free parameters; 
    \item the corrected Akaike Information Criterion \citep[AICc,][]{Akaike1974, CAVANAUGH1997201}, defined as 
    \begin{equation}
        \mathrm{AICc} = 2  k + \chi^2 + \frac{2k(k+1)}{n-k-1} \, .
    \end{equation}
    \end{enumerate}
The BIC and AICc estimators penalize models with increasing number of free parameters, to contrast overfitting. Thus, models with lower BIC and AICc vales are preferred.

\subsection{Modeling with A, B}
\label{subs:pointAB}
At the beginning, we consider in the strong lensing modeling only the positions of the spectroscopically-confirmed eight multiple images of the sources A and B. 

First, we assume a pseudo-isothermal total mass distribution \citep[PIEMD;][]{Kneib1996} for the main lens. In {\tt GLEE}, it is described by six parameters: the $x$ and $y$ coordinates of the center, the semi-major ($a$) to semi-minor ($b$) axis ratio $q=b/a$, the position angle $\theta$, the Einstein radius $\theta_E$, and the core radius $r_{core}$. Here, we fix $r_{core}=0$. Note that the value of $\theta_E$ is defined for a source at $z = \infty$ and does not correspond to that of the Einstein radius of the system, which should be nearly independent of the mass modeling details. The value of $\theta_E$ is a parameter which describes the lens strength and enters the dimensionless surface mass density $\kappa_\mathrm{PIEMD}$ as
\begin{equation}
    \kappa_\mathrm{PIEMD} (x,y) = \frac{\theta_E}{2 \sqrt{r_{core}^2+\frac{x^2}{(1+e)^2}+\frac{y^2}{(1-e)^2}}} \, ,
\end{equation}
where the ellipticity $e=\frac{1+q}{1-q}$. The value of the physical Einstein radius, as well as being estimated from this equation, would be clearly estimated from the total mass profiles presented in the following.

Then, we assume a singular power law elliptical mass distribution \citep[SPEMD;][]{Barkana1998}. In {\tt GLEE}, it is described by seven parameters: the first six are in common with the PIEMD profile, and the slope $g$, which is related to the three-dimensional logarithmic density slope $\gamma'= \diff \log[\rho(r)]/\diff \log(r)$ (i.e., $\rho \propto r^{-\gamma'}$) through $\gamma'=2 g +1$ (i.e., an isothermal profile corresponds to $\gamma'=2$ and $g = 0.5$). In the following, we will refer to the physical parameter $\gamma'$.
Similarly to the PIEMD case, $\theta_E$ is a parameter of the mass distribution introduced in the dimensionless surface mass density $\kappa_\mathrm{SPEMD}$ as
\begin{equation}
    \kappa_\mathrm{SPEMD}(x,y) = \theta_E \left( x^2+\frac{y^2}{q^2} + \frac{4r^2_{core}}{(1+q)^2}\right) ^ {-\frac{\gamma^\prime-1}{2}}.
    \end{equation}

In both the mass distributions, we include an external shear component. In {\tt GLEE}, this is described by two parameters: the shear strength, $\gamma_\mathrm{ext}$, and its position angle, $\phi_\mathrm{ext}$ (where $0$ means that images are stretched horizontally, along the $x$-axis, and $\pi/2$ means images are stretched vertically, along the $y$-axis).

The best-fit values of the parameters for these first two models are reported in the two upper entries of Table~\ref{tab:point models}. The values of the $x$ and $y$ coordinates are in arcseconds, relative to the center of light of the elliptical lens galaxy, i.e., its brightest pixel. For each model, we list the best-fit parameter values, the number of observables ($N_{obs}$), the number of degrees of freedom (dof), and the value of the minimum ($\chi^2_{min}$).

\subsection{Modeling with A, B, C}
\label{subs:pointABC}
Then, we add the two multiple images of source C, and use the same two mass density profiles defined in the previous section to fit the positions of all ten multiple images. The introduction of C requires an additional free parameter, i.e. its unknown redshift. In {\tt GLEE}, this is parametrized with the value of $D_{ds}/D_s$, which is the ratio between the distances between the deflector and the source and between the observer and the source. Below, we will convert the values (and the posterior probability distribution) of $D_{ds}/D_s$ into those corresponding to the redshift of the source, $z_C$, and we will always refer to this parameter. The best-fit parameter values are reported in the third and fourth entries of Table \ref{tab:point models}.

Since the position of the multiple image C2 is angularly very close to the main lens galaxy, we decide to try also PIEMD and SPEMD mass density profiles with the value of $r_{core}$ free to vary. We will refer to these models as PIEMD$+$rc and SPEMD$+$rc. We expect that C2 can affect the reconstructed lens total mass distribution in the inner region, depending mainly on the values of $r_{core}$ and $\gamma'$. In fact, from strong lensing data, one can accurately measure the total mass distribution of a lens at the radii where the multiple images are observed. The presence of the source C allows us to expand the radial interval from approximately (30--50) to (15--63)~kpc. 
The best-fit values of the parameters of the models  PIEMD$+$rc and SPEMD$+$rc are reported in the two lower entries of Table \ref{tab:point models}.

By comparing the best-fit parameter values of these different models, we notice that the center of the total mass is always shifted with respect to the luminosity center, along the positive $x$-axis and the negative $y$-axis. Furthermore, the values of the angles $\theta$ and $\phi_\mathrm{ext}$ are similar and approximately orthogonal in almost all the models. When the source C is included, the total mass of the deflector becomes more elliptical (i.e., the value of $q$ decreases), unless a non-vanishing core radius is considered. Moreover, the introduction of source C and of a possible core radius changes significantly the best-fit values of the Einstein radii, suggesting a degeneracy between the values of the $\theta_E$, $r_{core}$, $\gamma'$, and $z_C$ parameters. Finally, the best-fit SPEMD profiles without a core radius are shallower than the corresponding isothermal profiles ($\gamma' \approx 1.5$ vs $\gamma'=2$), while the best-fit values of core radius and slope are very similar, when the former is allowed to vary.

\begin{figure}
   \centering
   \includegraphics[width=8cm]{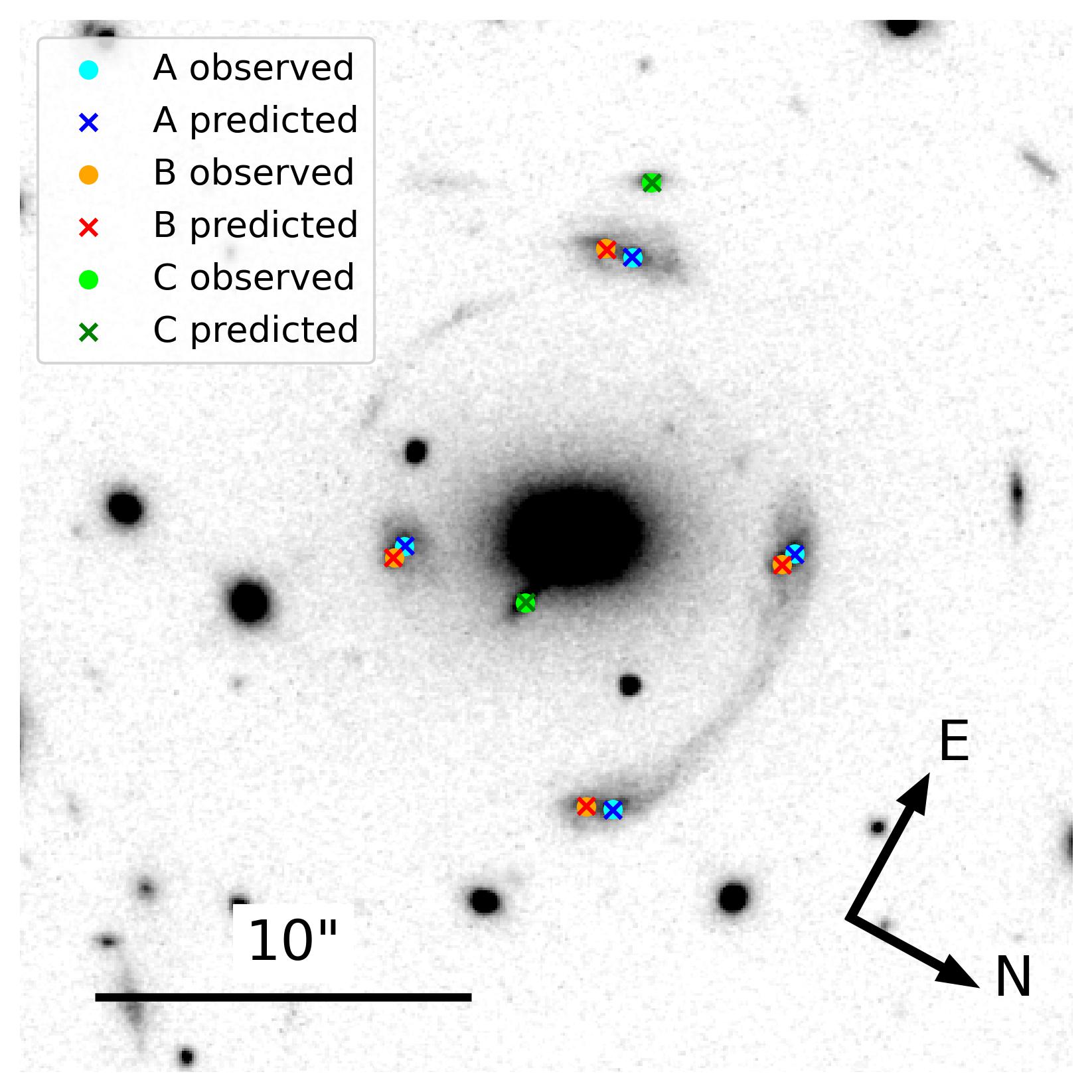}
      \caption{\textit{HST}/F160W image of SDSS\,J0100+1818 and comparison between the observed multiple images of the sources A (blue), B (red), and C (green), represented with circles, and those predicted by the best-fit PIEMD$+$rc model (fifth entry of Tab.~\ref{tab:point models}), represented with crosses with the corresponding colors.}
         \label{fig:comparison}
   \end{figure}

\begin{table}[]
\centering
\scriptsize
\begin{tabular}{@{}lccccccc@{}}
\toprule
\toprule
\multicolumn{3}{l}{Multiple images A, B}  & & & & & \\
PIEMD & & & & & & &  \\ 
 $x$ [$''$] & $y$ [$''$] & $q$ & $\theta$ [rad] & $\theta_E$ [$''$] &  $r_{core}$ [$''$]  &  \\ 
 0.27 & $-0.13$ & 0.69 & 0.07 & 11.1 & [0.0] & \\
shear & $\gamma_\mathrm{ext}$ & $\phi_\mathrm{ext}$ [rad] & & & & & \\
& 0.21 & 1.44 & & & & & \\
& \multicolumn{2}{c}{$N_{obs}=16$} & \multicolumn{2}{c}{dof $=5$} & \multicolumn{2}{c}{$\chi^2_{min} = 1.29$} & \\ 
& \multicolumn{2}{c}{$\mathrm{BIC}=31.8$} & \multicolumn{2}{c}{$\mathrm{AICc}=89.3$} & \multicolumn{2}{c}{$\mathrm{rms}=0.028\arcsec$} & \\ \midrule
SPEMD & & & & & & & \\
$x$ [$''$] & $y$ [$''$] & $q$ & $\theta$ [rad] & $\theta_E$ [$''$] &  $r_{core}$ [$''$]  & $\gamma'$ \\
0.25 & $-0.11$ & $0.77$ & $0.10$ & $3.8$ &  [0.0]  & 1.55 \\ 
shear & $\gamma_\mathrm{ext}$ & $\phi_\mathrm{ext}$ [rad] & & & & & \\
 & $0.06$ & $1.08$ & & & & & \\
& \multicolumn{2}{c}{$N_{obs}=16$} & \multicolumn{2}{c}{dof $=4$} & \multicolumn{2}{c}{$\chi^2_{min} = 0.63$} & \\
& \multicolumn{2}{c}{$\mathrm{BIC}=33.9$} & \multicolumn{2}{c}{$\mathrm{AICc}=128.6$} & \multicolumn{2}{c}{$\mathrm{rms}=0.020\arcsec$} & \\ \midrule
\multicolumn{3}{l}{Multiple images A, B, C}  & & & & & \\

PIEMD & & & & & & &  \\ 
$x$ [$''$] & $y$ [$''$] & $q$ & $\theta$ [rad] & $\theta_E$ [$''$] &  $r_{core}$ [$''$]  &  \\  
0.17 & $-0.22$ & $0.41$ & $0.01$ & $12.9$ &  $[0.0]$  &  \\ 
shear & $\gamma_\mathrm{ext}$ & $\phi_\mathrm{ext}$ [rad] & & & & & \\
 & $0.06$ & 1.22 & & & & & \\
source & \multicolumn{2}{c}{$D_{ds}/D_s = 0.477$} & \multicolumn{2}{c}{(corresponding to $z_C = 1.37$)} & &\\
& \multicolumn{2}{c}{$N_{obs}=20$} & \multicolumn{2}{c}{dof $=6$} & \multicolumn{2}{c}{$\chi^2_{min} = 4.04$} & \\ 
& \multicolumn{2}{c}{$\mathrm{BIC}=46.0$} & \multicolumn{2}{c}{$\mathrm{AICc}=116.0$} & \multicolumn{2}{c}{$\mathrm{rms}=0.044\arcsec$} & \\ \midrule
SPEMD & & & & & & & \\
$x$ [$''$] & $y$ [$''$] & $q$ & $\theta$ [rad] & $\theta_E$ [$''$] &  $r_{core}$ [$''$]  & $\gamma'$ \\
 $0.14$ & $-0.14$ & $0.60$ & 0.04 & $4.0$ &  [0.0]  & $1.53$ \\ 
shear & $\gamma_\mathrm{ext}$ & $\phi_\mathrm{ext}$ [rad] & & & & & \\
 & 0.12 & $0.17$ & & & & & \\
source & \multicolumn{2}{c}{$D_{ds}/D_s = 0.548 $} & \multicolumn{2}{c}{(corresponding to $z_C = 1.72$)} & &\\
& \multicolumn{2}{c}{$N_{obs}=20$} & \multicolumn{2}{c}{dof $=5$} & \multicolumn{2}{c}{$\chi^2_{min} = 1.78$} & \\ 
& \multicolumn{2}{c}{$\mathrm{BIC}=46.7$} & \multicolumn{2}{c}{$\mathrm{AICc}=151.8$} & \multicolumn{2}{c}{$\mathrm{rms}=0.030\arcsec$} & \\ \midrule
PIEMD$+$rc & & & & & & &  \\ 
$x$ [$''$] & $y$ [$''$] & $q$ & $\theta$ [rad] & $\theta_E$ [$''$] &  $r_{core}$ [$''$]  &  \\  
0.18 & $-0.09$ & $0.81$ & $0.12$ & $17.2$ &  $2.8$  &  \\ 
shear & $\gamma_\mathrm{ext}$ & $\phi_\mathrm{ext}$ [rad] & & & & & \\
 & $0.10$ & 1.32 & & & & & \\
source & \multicolumn{2}{c}{$D_{ds}/D_s = 0.583$} & \multicolumn{2}{c}{(corresponding to $z_C = 1.98$)} & &\\
& \multicolumn{2}{c}{$N_{obs}=20$} & \multicolumn{2}{c}{dof $=5$} & \multicolumn{2}{c}{$\chi^2_{min} = 1.08$} & \\ 
& \multicolumn{2}{c}{$\mathrm{BIC}=46.0$} & \multicolumn{2}{c}{$\mathrm{AICc}=151.1$} & \multicolumn{2}{c}{$\mathrm{rms}=0.023\arcsec$} & \\ \midrule
SPEMD$+$rc & & & & & & & \\
$x$ [$''$] & $y$ [$''$] & $q$ & $\theta$ [rad] & $\theta_E$ [$''$] &  $r_{core}$ [$''$]  & $\gamma'$ \\
 $0.18$ & $-0.11$ & $0.78$ & 0.09 & $10.9$ &  2.5  & $2.10$ \\ 
shear & $\gamma_\mathrm{ext}$ & $\phi_\mathrm{ext}$ [rad] & & & & & \\
 & 0.11 & $1.36$ & & & & & \\
source & \multicolumn{2}{c}{$D_{ds}/D_s = 0.577 $} & \multicolumn{2}{c}{(corresponding to $z_C = 1.93$)} & &\\
& \multicolumn{2}{c}{$N_{obs}=20$} & \multicolumn{2}{c}{dof $=4$} & \multicolumn{2}{c}{$\chi^2_{min} = 0.62$} & \\ 
& \multicolumn{2}{c}{$\mathrm{BIC}=48.6$} & \multicolumn{2}{c}{$\mathrm{AICc}=213.9$} & \multicolumn{2}{c}{$\mathrm{rms}=0.017\arcsec$} & \\
\bottomrule
\end{tabular}
\caption{Modeling of the deflector total mass distribution using the observed multiple image positions. The two upper models were optimized by exploiting the eight observed positions of the spectroscopically confirmed background sources A and B, while the four bottom models include also the observed positions of the two multiple images of source C, without a spectroscopic redshift measurement. For each model, we show the best-fit values of the parameters, the number of degrees of freedom (dof), the minimum chi-square ($\chi^2_{min}$) value, and the resulting BIC, AICc and rms values. The values of the $x$ and $y$ coordinates are referred to the center of light of the main elliptical lens galaxy, i.e., to its brightest pixel. The position angle $\theta$ is measured counter clockwise from $x$. Values in square brakets are kept fixed. }
\label{tab:point models}
\end{table}
The low values of the minimum $\chi^2$ suggest that we can accurately reproduce the positions of the observed multiple images by assuming a one-component total mass distribution for the deflector. In fact, the average distance between the observed and model-predicted positions of the multiple images is only of $0.02 \arcsec$. 
In Fig.~\ref{fig:comparison}, we show the observed (filled circles) and the model-predicted (crosses) positions of the multiple images for the PIEMD$+$rc model. All the considered models would result in very similar figures.    

Furthermore, on the top of Table \ref{tab:mcmc} we report the median and the $68\%$ confidence level uncertainty values of the parameters of the different models. They are extracted from Markov chain Monte Carlo (MCMC) posteriors with $10^6$ steps, with acceptance rates between $20\%$ and $30\%$, after excluding the first $10\%$ steps (considered as the burn-in phase). To correctly compare different models with different numbers of observables and degrees of freedom, for each model we have rescaled the errors on the observed multiple images so that the value of  $\chi^2_{min}$ is approximately equal to that of $\mathrm{dof}$. 

\begin{figure}[h!]
   \centering
   \includegraphics[width=0.5\textwidth]{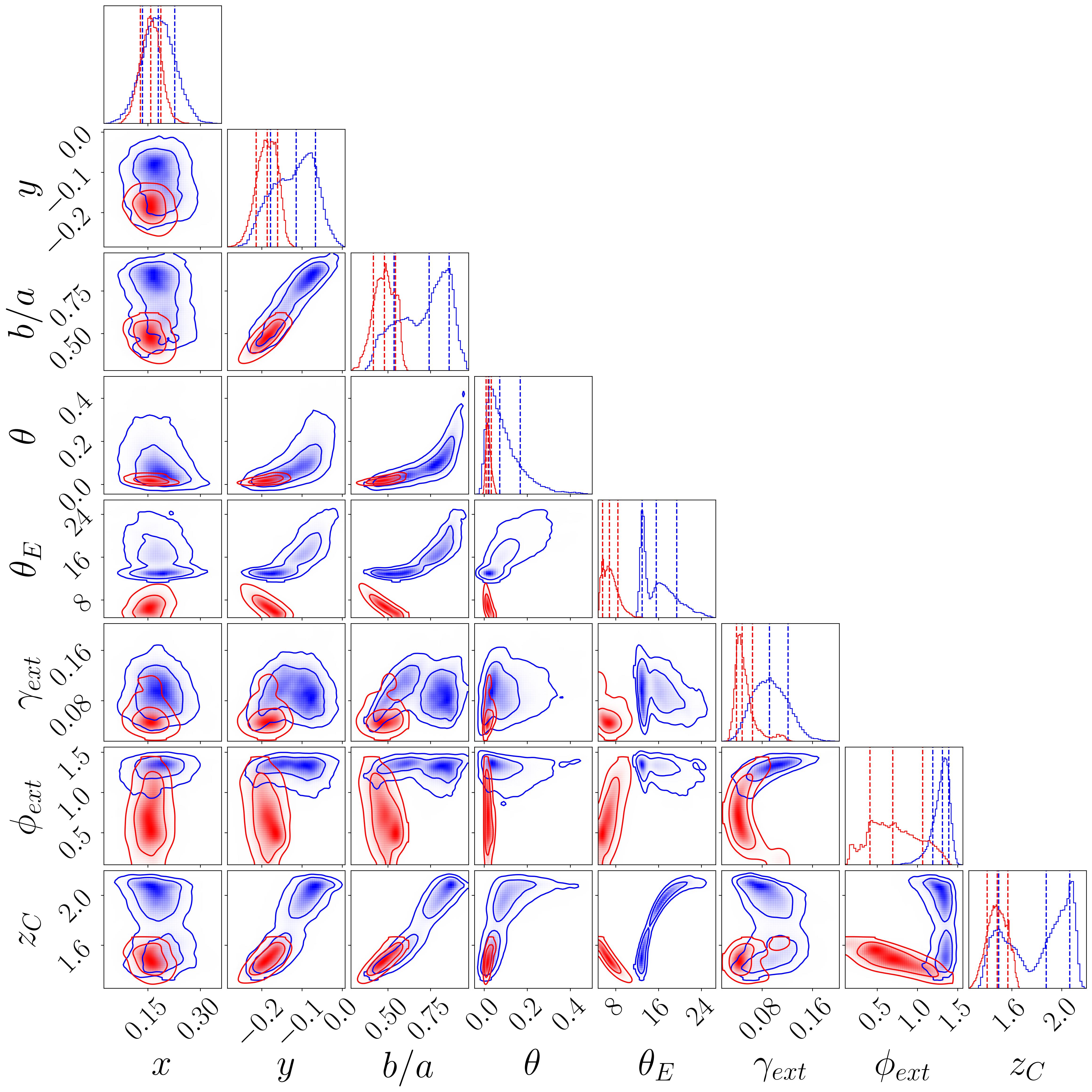}
   \caption{Probability density distributions of the parameters of the PIEMD$+$rc (in blue) and SPEMD (in red) models. The marginalized 1D histograms of each parameter are shown along the diagonal, while the other panels show the joint 2D probability distributions of the two parameters reported on the horizontal and vertical axes. The parameters plotted are those introduced in Sect.~\ref{subs:pointAB}. The vertical dashed lines in the 1D histograms represent the $16^\mathrm{th}$, $50^\mathrm{th}$ and $84^\mathrm{th}$ percentiles, while the solid lines in the 2D distributions represent the $0.68$ and $0.95$ contour levels.  }
              \label{fig:corner_models}%
    \end{figure}

In Fig.~\ref{fig:corner_models}, we show the posterior probability density distributions of the parameters of the PIEMD$+$rc (in blue) and SPEMD (in red) models, with sources A, B and C included in the modeling. In order to compare them, only the common parameters are plotted, and thus $\gamma^\prime$ and $r_{core}$ are not included. The $x$, $y$, $q$ and $\theta$ distributions of the PIEMD$+$rc and SPEMD models are consistent. The Einstein radius distributions are centered on different values, but this is not surprising: its value is degenerate with those of the slope $\gamma^\prime$ (see Fig.~\ref{fig:corner_3}) and of the core radius $r_{core}$. Finally, we remark that the SPEMD model shows a unimodal probability density distribution for $z_C$, as opposed to the bimodal one of the PIEMD$+$rc model.  

\begin{figure}
   \centering
   \includegraphics[width=0.5\textwidth]{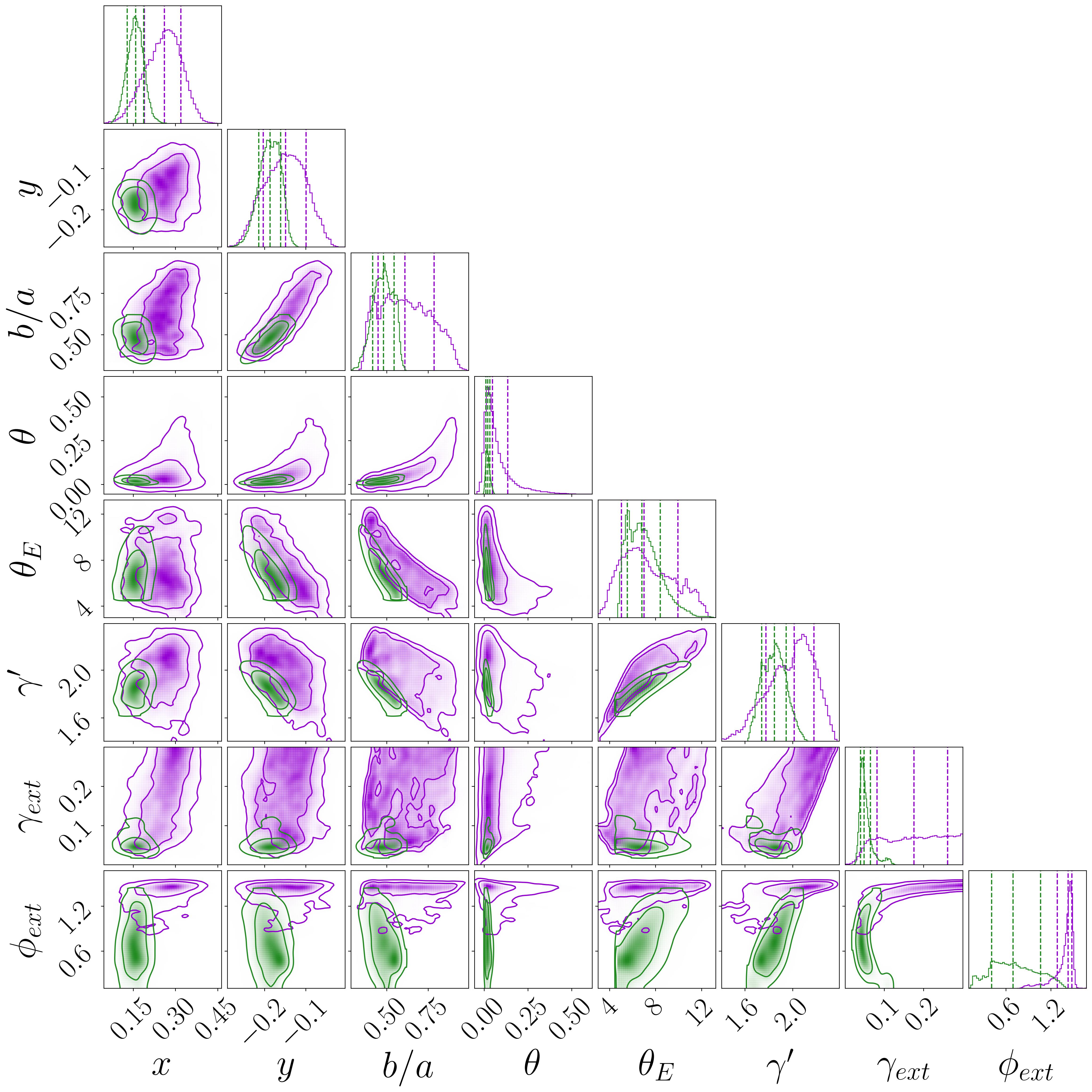}
   \caption{Probability density distribution of the parameters of the SPEMD model considering the multiple image positions of sources A and B only (in purple) and of the sources A, B and C (in green). 
   The panels show the probability distributions as described in Fig.~\ref{fig:corner_models}.}
              \label{fig:corner_2vs3_src}%
    \end{figure}

Finally, we show in Fig.~\ref{fig:corner_2vs3_src} the comparison between two models with the same total mass parameterization (i.e., SPEMD), optimized first with the eight multiple images of sources A and B (in purple) and then with the ten multiple images of sources A, B and C (in green). We can conclude that the results obtained with only two sources are consistent with those obtained with three sources and that the introduction of source C, at a different redshift, significantly reduces all but one confidence intervals. 
   
\subsection{Total mass profile of the deflector}   
\label{subs:total_mass_point}
For each model, we estimate the total mass ($M_T$) distribution of the deflector by extracting randomly $1000$ models from the $10^6$ steps of the last MCMCs described above. First, we convert the convergence maps provided by {\tt GLEE} into the corresponding total mass maps. Then, we sum up the contribution of all the pixels within circular apertures centered on the brightest pixel of the main lens galaxy, with a step of 0.5 pixels, to obtain the cumulative total mass profiles that are presented in the following. Within each fixed aperture, we consider the distribution of the total mass values of all the $1000$ random models, from which we measure the median value and the $16^\mathrm{th}$ and $84^\mathrm{th}$ percentiles, as the uncertainties at the $1\sigma$ confidence level.

We start comparing the cumulative total mass profiles of each different mass model reconstructed by using the multiple images of sources A and B only and A, B and C. We find that the profiles show consistent median values and that the introduction of source C significantly reduces the statistical uncertainties in the inner and outer regions. For example, for the SPEMD model, when source C is considered, the relative uncertainties are reduced from about $32\%$ to $17\%$ and from $9\%$ to $5\%$ in the inner ($R \approx 15$ kpc) and outer ($R \approx 63$ kpc) regions, respectively. Thus, in the following discussion and figures, we will refer only to the models optimised by taking into account the observed multiple image positions of the sources A, B, and C. 

\begin{figure}[!h]
   \centering
   \includegraphics[width=0.48\textwidth]{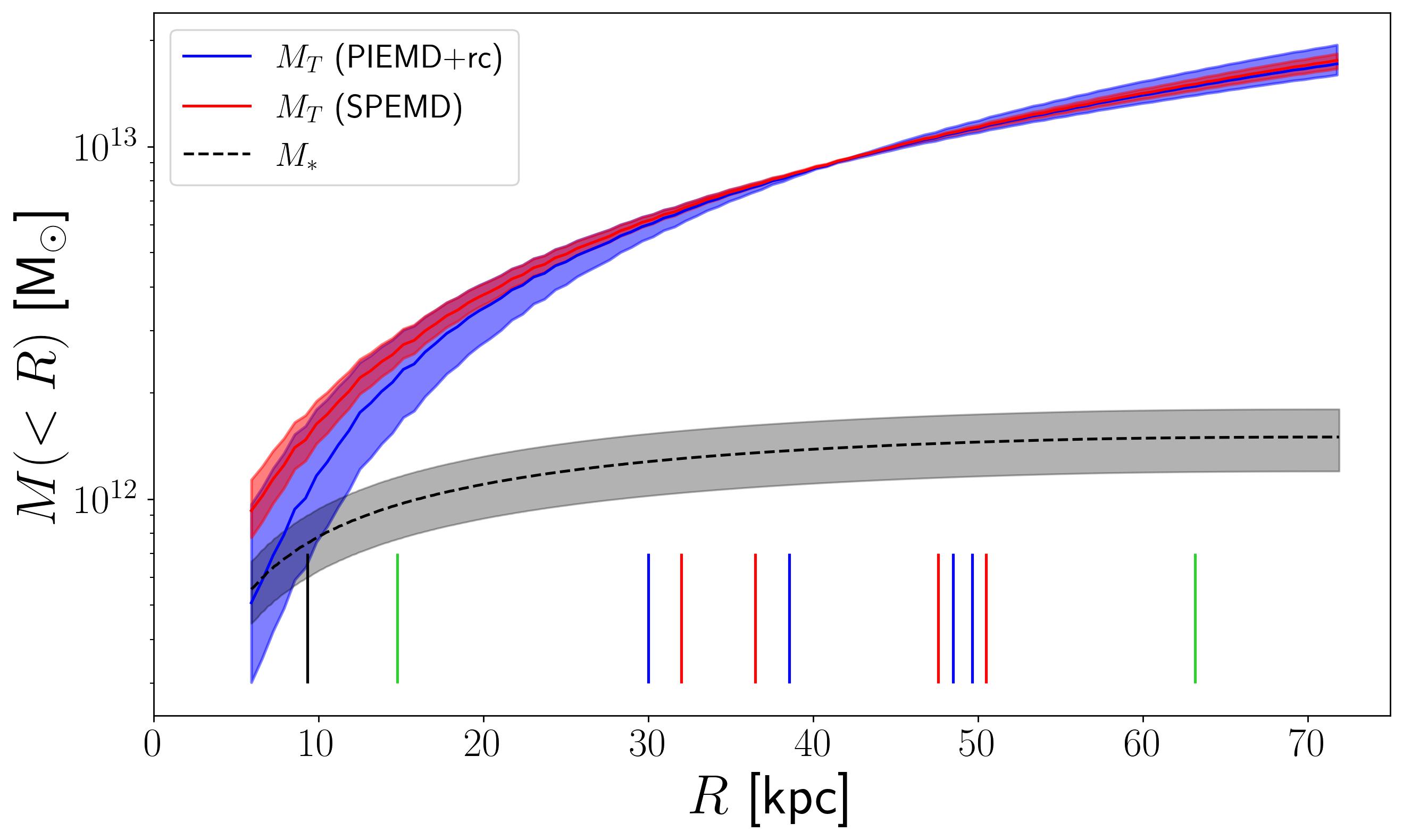}
      \caption{Cumulative projected total mass profiles for the PIEMD+rc (blue) and SPEMD (red) models with $\pm 1\sigma$ uncertainties (shaded areas), obtained by modeling the multiple images of A, B and C as point-like objects. In dashed black, the cumulative projected stellar mass profile from Fig.~\ref{fig:stellarmass_profile} is shown. The vertical lines close to the $x$-axis locate the distances from the lens galaxy center of the different multiple images, color-coded following Figs.~\ref{fig:labels} and \ref{fig:comparison}. The black line shows the effective radius of the main lens galaxy.}
         \label{fig:comparison_piemd-spemd}
   \end{figure}

In Fig.~\ref{fig:comparison_piemd-spemd}, we show the cumulative projected total mass profiles of the deflector for the PIEMD$+$rc and SPEMD models. Despite the different total mass density parameterizations, and the slightly different optimized values of the parameters, the profiles are very similar and agree on the same mass value, with the smallest statistical errors, at the mean distance ($R \approx 42$ kpc) of the multiple images from the lens center, which represents approximately the physical Einstein radius of the system. The relative uncertainties at the distances of the innermost (C2, $R \approx 15$ kpc) and outermost (C1, $R \approx 63$ kpc) multiple images are, respectively, about $17\%$ and $5\%$ for the SPEMD model, while its minimum of approximately $1\%$ is reached at $R \approx 42$ kpc. The total projected mass value measured within this radius is of $(9.06 \pm 0.04) \times 10^{12}$ M$_\odot$ for the PIEMD$+$rc model and of $(9.09 \pm 0.05) \times 10^{12}$ M$_\odot$ with the SPEMD model.  
In the same figure, we also include the stellar mass profile of the system we described above.

\begin{figure}[!h]
   \centering
   \includegraphics[width=0.48\textwidth]{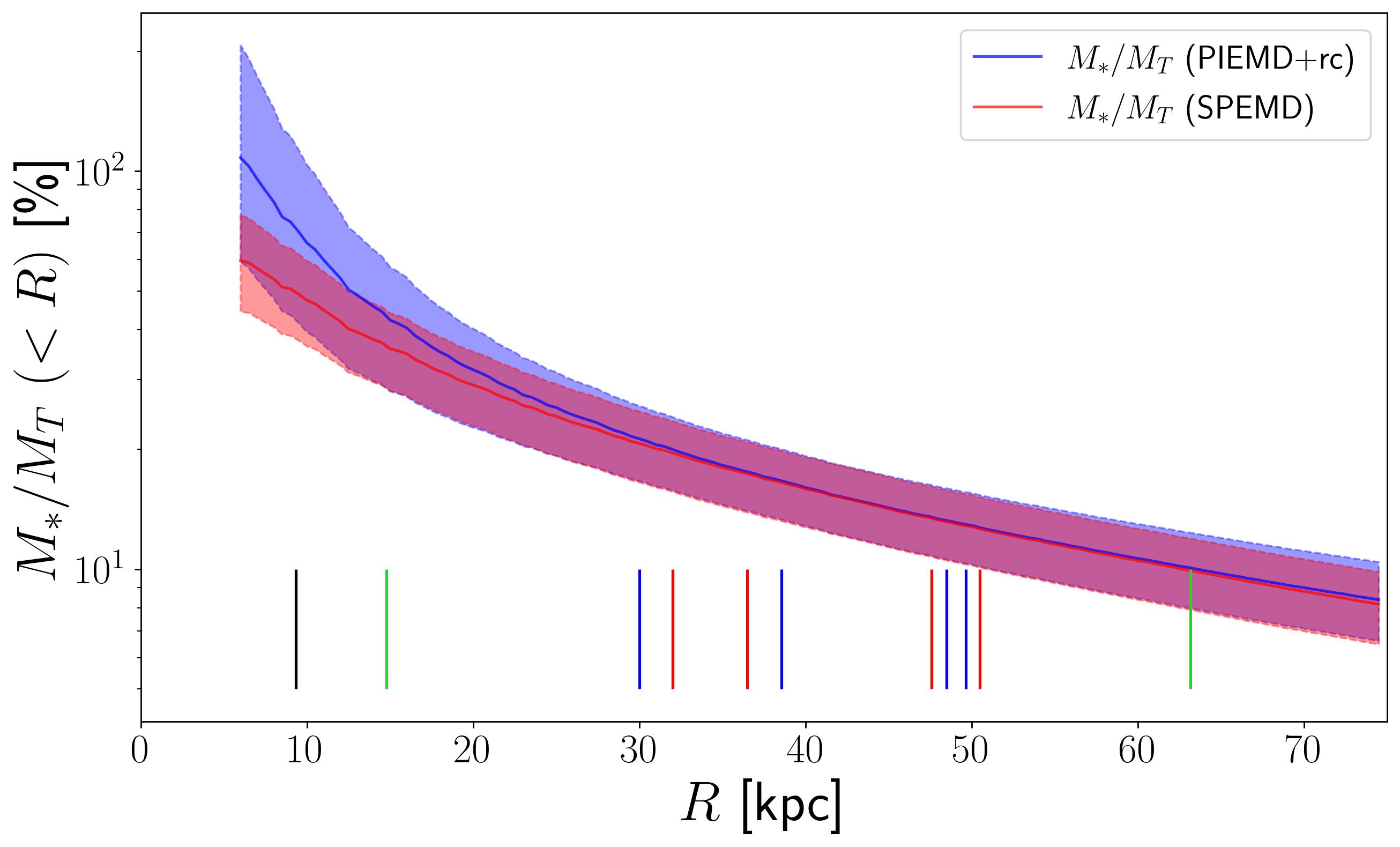}
      \caption{Cumulative projected stellar-over-total mass fraction profiles for the PIEMD+rc (blue) and SPEMD (red) models with $\pm 1\sigma$ uncertainties (shaded areas), obtained by modeling the multiple images of A, B and C as point-like objects. The vertical lines are the same as in Fig.~\ref{fig:comparison_piemd-spemd}.}
         \label{fig:ratio_model}
   \end{figure}

Finally, in Fig.~\ref{fig:ratio_model}, we plot the cumulative projected stellar-over-total mass fraction profiles for the PIEMD$+$rc and SPEMD models. The two profiles are consistent, given the uncertainties, and differ mainly in the inner regions. At the lens galaxy effective radius,  we estimate values of $(70 \pm 29)\%$ and $(49 \pm 12)\%$ for the PIEMD$+$rc and SPEMD models, respectively. According to both models, the fraction values are of $(15 \pm 3)\%$ and $(10 \pm 2)\%$ at $R \approx 42$ kpc and $R \approx 63$ kpc, respectively. 

\section{Extended source modeling}
\label{sec:extended}
In this section, we describe the reconstruction of the extended surface brightness (SB) of the lensed sources that we performed with {\tt GLEE}. At this stage, we merge the sources A and B into a single, double-peaked source, at $z=1.88$. Thus, in the following, we will refer to \textit{single source} models as the analogous of modeling with A and B in the point-like source modeling, and to \textit{two source} models as of modeling with A, B, and C.

To model the SB of extended images, {\tt GLEE} needs as input a mask with the pixels containing the observed  multiple images and possible arcs of the lensed sources. These pixels are mapped onto the source plane through the ray-tracing equation, where the source SB is reconstructed on a regular grid of pixels. The ray-tracing equation depends on the total mass distribution of the lens and on the angular diameter distances between the lens and the sources. When we model the observed images of a single source (i.e., A and B), with its spectroscopic redshift value measured and considered fixed, {\tt GLEE} finds the best-fit model by varying the values of the parameters of the mass distribution of the lens, and optimizing the posterior probability distribution, which is proportional to the product of the likelihood and the prior of the lens mass parameters \citep{Suyu2010}. Instead, when we model the images of the two sources, the mapping depends also on the relative distances between the observer, the deflector, and source C, thus its redshift is introduced as an additional free parameter.

In the optimization process, the pixelated SB distributions of the sources are mapped back onto the image plane to obtain the model-predicted images, with flux intensity $f_{ij}^\mathrm{pred}$, and compared with those observed, $f_{ij}^\mathrm{obs}$. The $\chi^2_\mathrm{SB}$ is thus defined as 
\begin{equation}
    \chi^2_\mathrm{SB} = \sum_{i,j} \frac{\left( f_{ij}^\mathrm{obs}-f_{ij}^\mathrm{pred} \right) ^2}{\sigma^2_{ij}} \, ,
\end{equation}
where $i$ and $j$ identify the position of each pixel. During the optimization, we impose a curvature regularization on the values of the source SB pixels, weighted by the value of the parameter $\hat{\lambda}$, which is included in the $\chi^2_\mathrm{SB}$ evaluation, as described in \cite{Suyu2006}. This parameter penalizes physically less plausible solutions with very irregular SB distributions for the sources. The number of degrees of freedom is here computed as 
\begin{equation}
    \mathrm{dof} = n_\mathrm{mask} - n_\mathrm{lens} - n_\mathrm{source} \, , 
\end{equation}
where $n_\mathrm{mask}$ is the number of pixels on the image plane included in the mask as observables, $n_\mathrm{lens}$ is the number of free parameters related to the total mass distribution of the lens, and $n_\mathrm{source}$ is the effective number (i.e., smaller than the adopted one) of pixels onto which the source is reconstructed on the source plane. At the beginning, we consider the SB of a source on a regular grid of $25 \times 25$ pixels. Then, we refine (and show here) the results adopting a $40 \times 40$ grid on the source plane. Note that {\tt GLEE}, after mapping the pixels included in the mask onto the source plane, reconstructs the source SB on a regular grid of pixels, with dimensions chosen by the user. Given that the maximum physical distance of the mapped pixels along the two coordinate axes is not known a priori, the SB source reconstructions usually have different pixel scales on the $x$ and $y$ axes, as can be observed in the plots presented in the following. 

\subsection{Modeling the deflector and the arcs}
\label{subs:ext_deflector}
We start from the best-fit models achieved with the point-like source modeling, described in Section \ref{sec:point}. In Table \ref{tab:point models}, we observe that the $\chi^2_{min}$ value decreases significantly, from approximately $4$ to $1$, when the number of degrees of freedom is reduced from $6$ to $5$ (from the PIEMD to the SPEMD and PIEMD$+$rc models). On the contrary, when dof is further reduced to $4$ (for the SPEMD$+$rc model), the value of $\chi^2_{min}$ shows only a minor decrement. Furthermore, the SPEMD$+$rc model has the largest values for both the BIC and AICc estimators. This suggests that this model is the least favored one, thus its extra complexity is statistically not fully justified. For these reasons, in the extended source modeling with two sources (AB and C), we decide to consider only the SPEMD and PIEMD$+$rc lens mass models, both with dof~$=5$ in the point-like source approximation.
The free parameters of each mass distribution are the same as described previously. In Table \ref{tab:extended model}, we report the best-fit values from the $\chi^2_\mathrm{SB}$ minimization with a $40 \times 40$ pixellated source. As before, we can observe some trends in the values of the reconstructed parameters. The center of the total mass distribution is slightly shifted from the light peak, along the positive $x$ and negative $y$ directions. The value of the magnitude of the external shear is high in the PIEMD model with a single source, but it significantly decreases when considering a SPEMD model or when the source C is introduced. The value of $\theta$ is approximately the same for all models, while the value of $q$ decreases for the SPEMD models, where the deflector is found to be more elliptical. Finally, both the SPEMD models suggest a lens total mass profile shallower than an isothermal one, and the value of $\gamma^\prime$ is almost identical, when a single or two background sources are considered. 

\begin{figure*}
   \centering
   \includegraphics[width=0.7\textwidth]{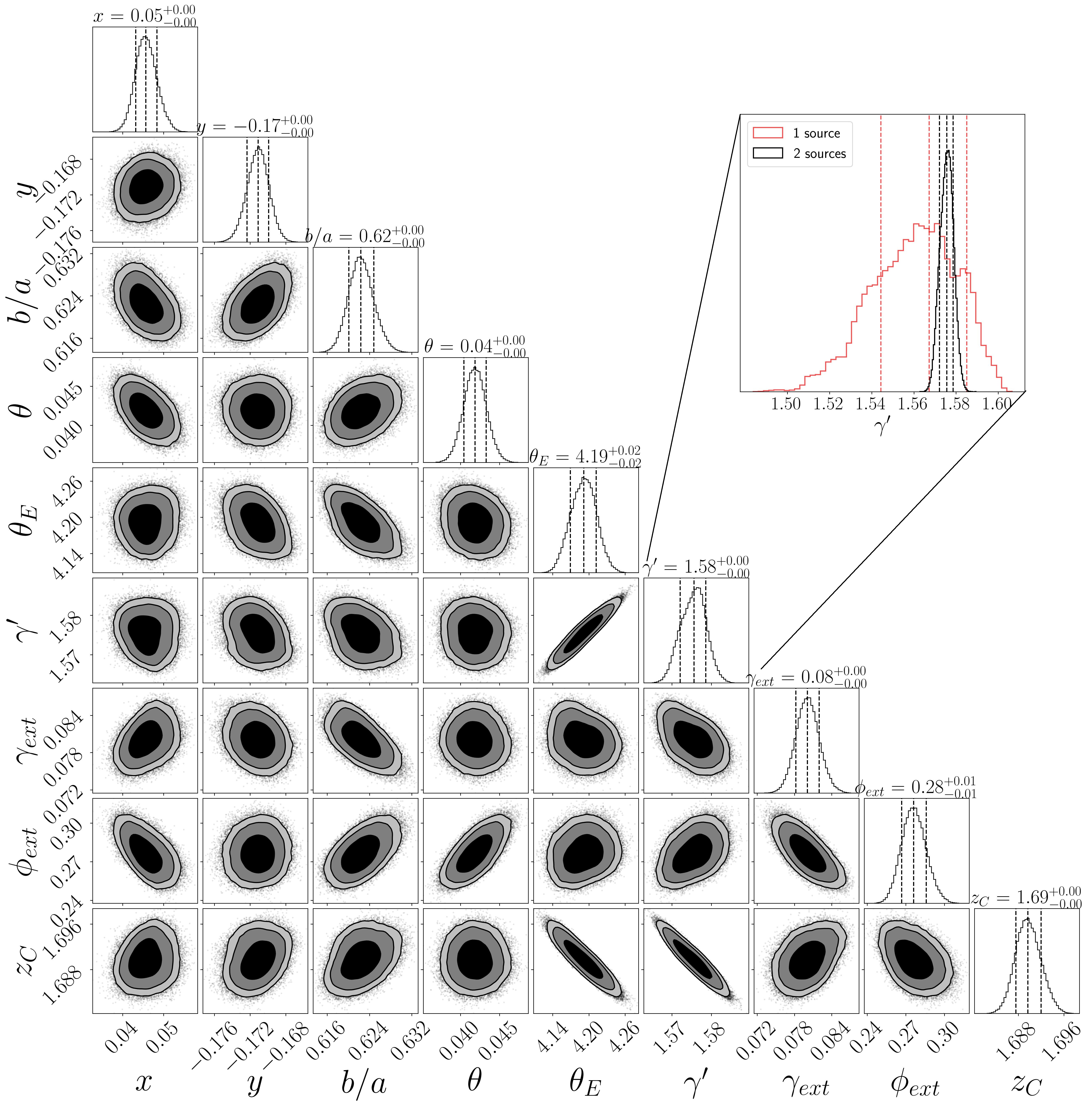}
   \caption{Probability density distribution for the SPEMD model with two sources (source AB and source C). The marginalized 1D histograms of each parameter are showed along the diagonal, while the other panels show the joint 2D probability distributions of the two parameters reported on the $x$ and $y$ axes. The parameters are those described in Sec.~\ref{subs:pointAB}. The vertical dashed lines in the 1D histograms represent the $16^\mathrm{th}$, $50^\mathrm{th}$ and $84^\mathrm{th}$ percentiles, which are also reported on top of each 1D histogram, while the solid lines and shaded areas in the 2D distributions represent the $0.68$, $0.95$, and $0.99$ contour levels. In the floating panel, we zoom in the marginalized 1D histogram of the $\gamma^\prime$ parameter (in black), and compare it with its distribution for the model with the AB source only (in red). The introduction of the source C provides tighter constraints on the value of the $\gamma^\prime$ parameter.}
              \label{fig:corner_3}%
    \end{figure*}

A comparison with the best-fit values of the point-like models shows that the center of the total mass distribution is always shifted in the same direction from the light distribution center. Moreover, the values of axis ratio, position angle, Einstein and core radii do not differ significantly when comparing each extended model with its corresponding point-like model. 
\begin{table}[]
\centering
\scriptsize
\begin{tabular}{@{}lccccccc@{}}
\toprule
\toprule
\multicolumn{3}{l}{Multiple images A, B}  & & & & & \\
PIEMD & & & & & & &  \\ 
$x$ [$''$] & $y$ [$''$] & $q$ & $\theta$ [rad] & $\theta_E$ [$''$] &  $r_{core}$ [$''$]  &  \\  
$0.24$ & $-0.14$ & $0.79$ & $0.14$ & $10.9$ &  [0.0]  &  \\ 
shear & $\gamma_\mathrm{ext}$ & $\phi_\mathrm{ext}$ [rad] & & & & & \\
 & $0.25$ & $1.45 $ & & & & & \\
& \multicolumn{2}{c}{$N_{obs}=5831$} & \multicolumn{2}{c}{dof $=4576$} & \multicolumn{2}{c}{$\chi^2_{\mathrm{SB,}min} = 2650.5$} & \\ \midrule
SPEMD & & & & & & & \\
$x$ [$''$] & $y$ [$''$] & $q$ & $\theta$ [rad] & $\theta_E$ [$''$] &  $r_{core}$ [$''$]  & $\gamma'$ \\ 
$0.18$ & $-0.14$ & $0.75$ & $0.09$ & $3.9$ &  [0.0]  & $1.57$ \\
shear & $\gamma_\mathrm{ext}$ & $\phi_\mathrm{ext}$ [rad] & & & & & \\
 & $0.06$ & $1.07$ & & & & & \\
& \multicolumn{2}{c}{$N_{obs}=5831$} & \multicolumn{2}{c}{dof $=4571$} & \multicolumn{2}{c}{$\chi^2_{\mathrm{SB,}min} = 2429.0$} & \\ \midrule \midrule
\multicolumn{3}{l}{Multiple images A, B, C}  & & & & & \\
SPEMD & & & & & & & \\
$x$ [$''$] & $y$ [$''$] & $q$ & $\theta$ [rad] & $\theta_E$ [$''$] &  $r_{core}$ [$''$]  & $\gamma'$ \\
$0.05$ & $-0.17$ & $0.62$ & $0.04$ & $4.2$ &  $[0.0]$  & $1.58$ \\
shear & $\gamma_\mathrm{ext}$ & $\phi_\mathrm{ext}$ [rad] & & & & & \\
 & $0.08$ & $0.28$ & & & & & \\
source & \multicolumn{2}{c}{$D_{ds}/D_s = 0.543$} & \multicolumn{2}{c}{(equals to $z_C = 1.69$)} & &\\
& \multicolumn{2}{c}{$N_{obs}=7195$} & \multicolumn{2}{c}{dof $=5396$} & \multicolumn{2}{c}{$\chi^2_{\mathrm{SB,}min} = 3188.3$} & \\ 
\midrule
PIEMD$+$rc & & & & & & &  \\ 
$x$ [$''$] & $y$ [$''$] & $q$ & $\theta$ [rad] & $\theta_E$ [$''$] &  $r_{core}$ [$''$]  &  \\ 
$0.09$ & $-0.12$ & $0.84$ & $0.15$ & $18.7$ &  $3.4$  &  \\
shear & $\gamma_\mathrm{ext}$ & $\phi_\mathrm{ext}$ [rad] & & & & & \\
 & $0.09$ & $1.25$ & & & & & \\
source & \multicolumn{2}{c}{$D_{ds}/D_s = 0.590$} & \multicolumn{2}{c}{(equals to $z_C = 2.04$)} & & \\
& \multicolumn{2}{c}{$N_{obs}=7195$} & \multicolumn{2}{c}{dof $=5583$} & \multicolumn{2}{c}{$\chi^2_{\mathrm{SB,}min} = 4400.9$} & \\ 
\bottomrule
\end{tabular}
\caption{Modeling of the deflector total mass distribution using the extended source modeling. The two upper models were optimized by exploting the double-peak source AB, while the two bottom models also include source C. The parameters are defined as in Tab.~\ref{tab:point models}.}
\label{tab:extended model}
\end{table}
The main differences for the lens reconstruction between the two ways of modeling the background sources can be observed in Table \ref{tab:mcmc}, where we list the median values and the statistical uncertainties extracted from Monte Carlo MCMC chains of $10^6$ steps, tuned to have $\chi^2_{min} \approx \mathrm{dof}$. The larger number of observables (approximately $7200$, instead of $20$) makes the marginalized probability distributions of the parameters much narrower when the sources are modeled as extended, as visible from the uncertainty values reported in the table. We show in Figure~\ref{fig:corner_3} the probability density distribution of the SPEMD model with two sources. There, we can observe the expected degeneracies between the values of $\theta_E$, $\gamma^\prime$ and $z_C$, found also in the point-like modeling. The values of the shear parameters show some significant degeneracies with those of several other parameters, like $q$ and $\theta$.  

\begin{figure*}
   \centering
   \includegraphics[width=4cm]{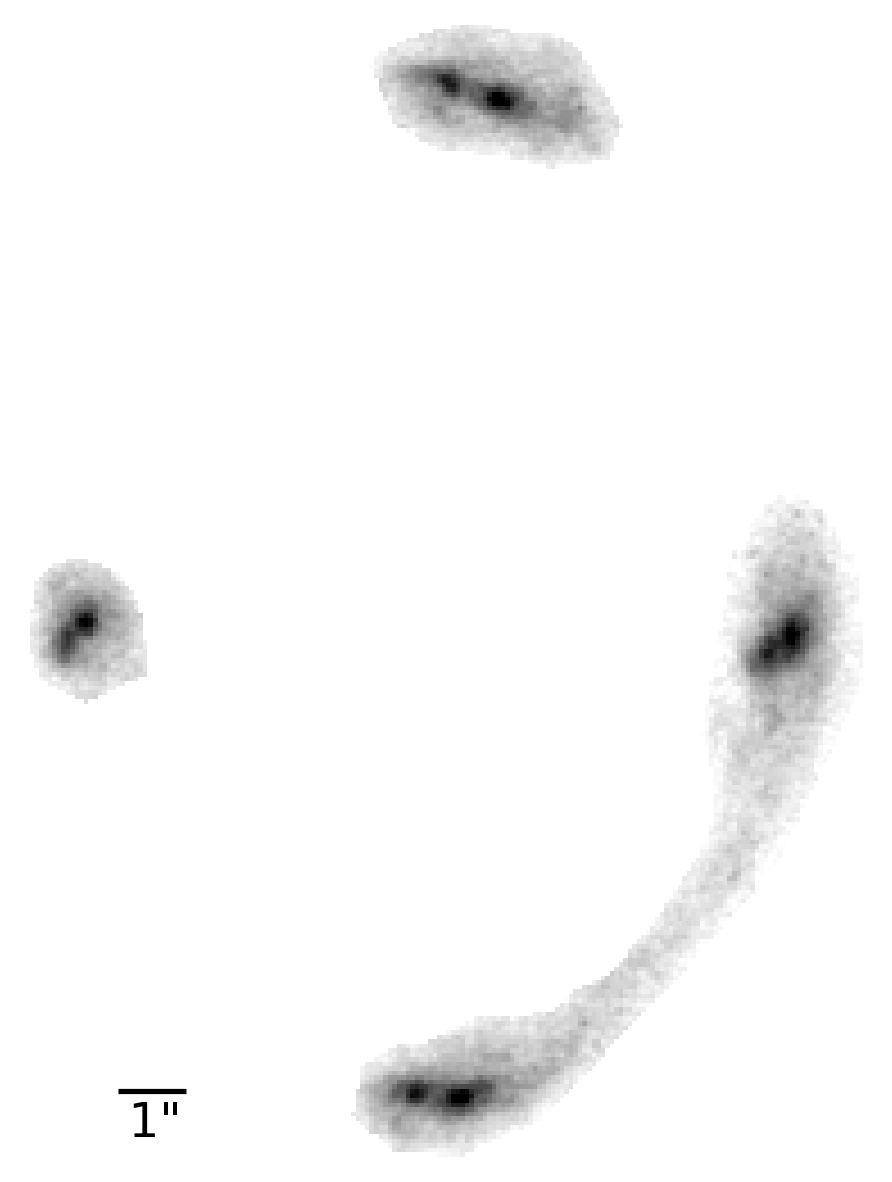}\hspace{2mm}
   \includegraphics[width=4cm]{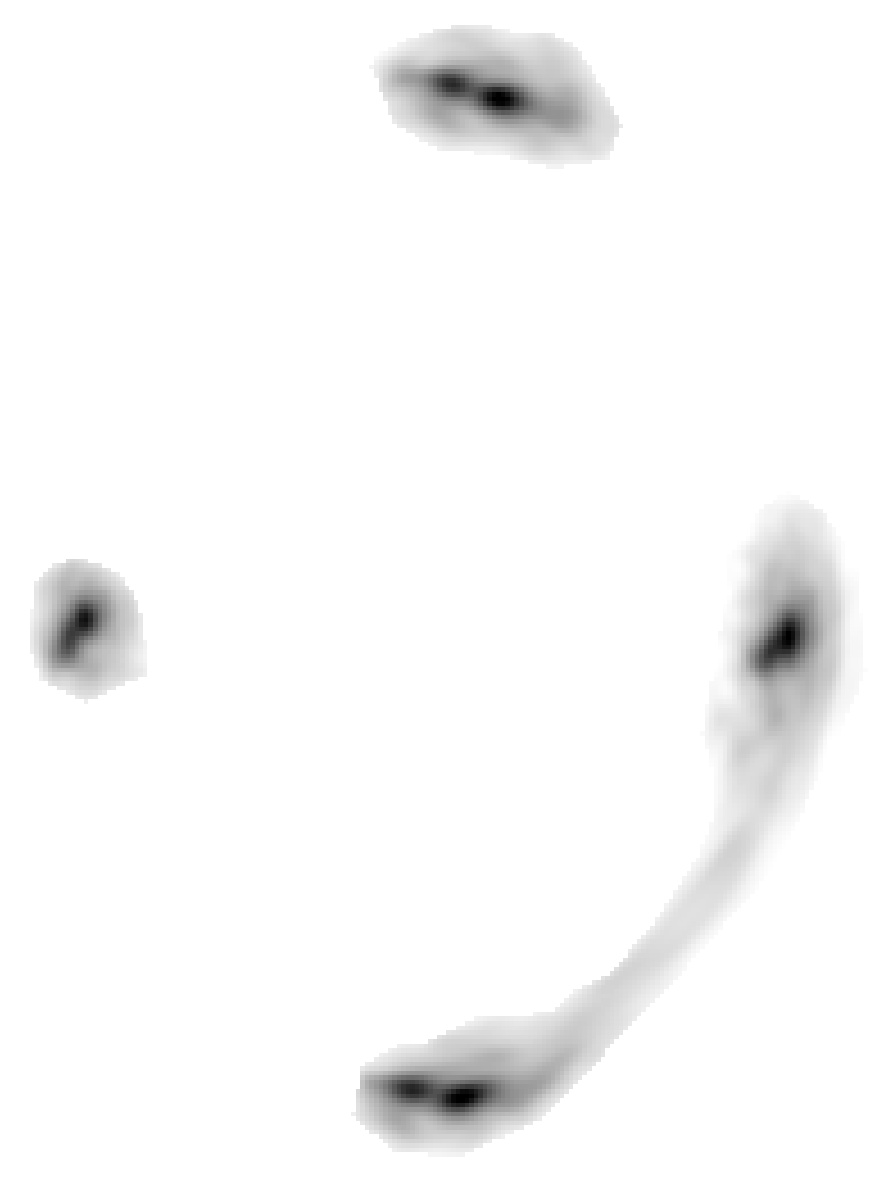}\hspace{2mm}
   \includegraphics[width=5cm]{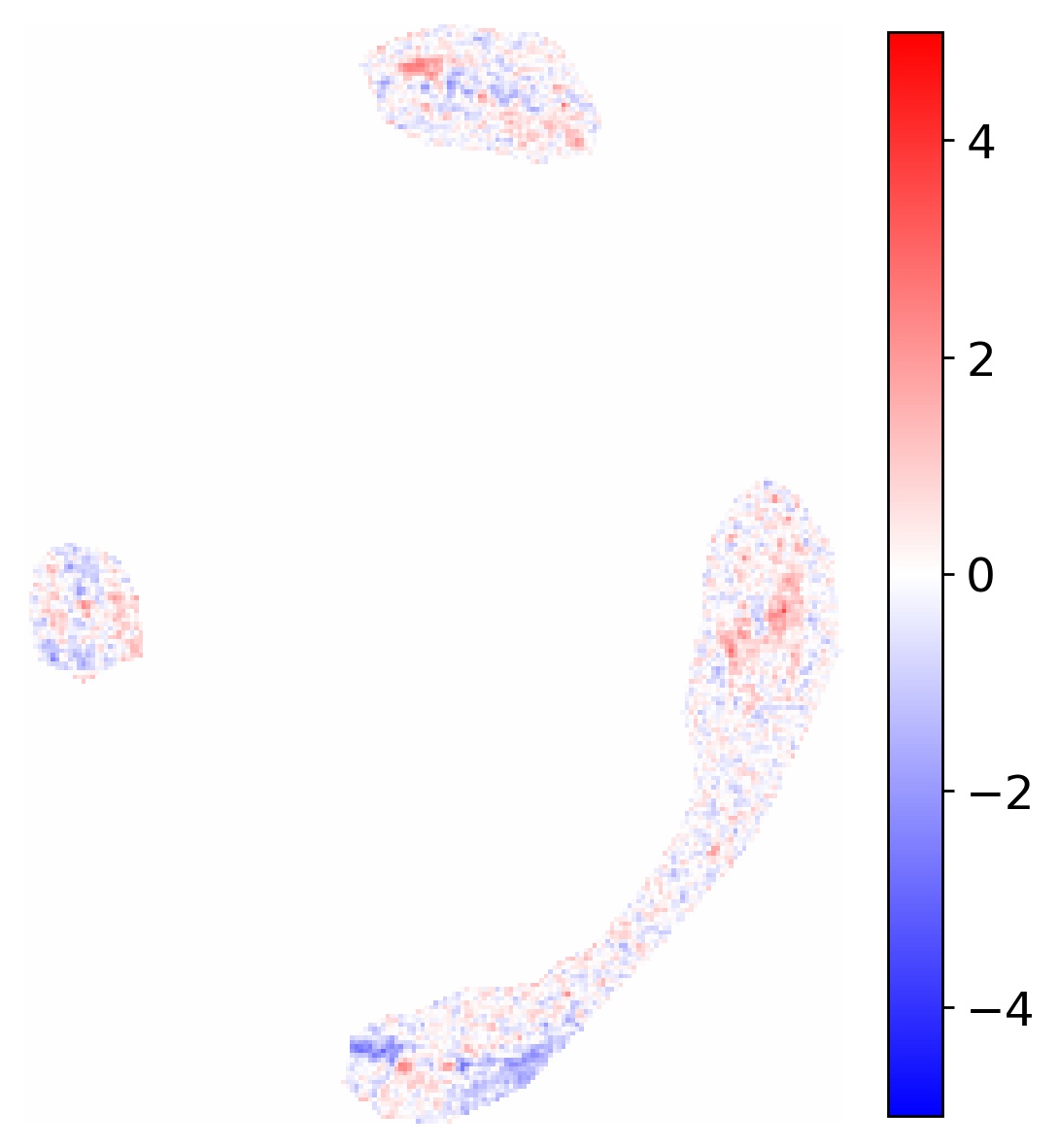}
   \hspace{5mm}
   \raisebox{0.25\height}{\includegraphics[width=3.5cm]{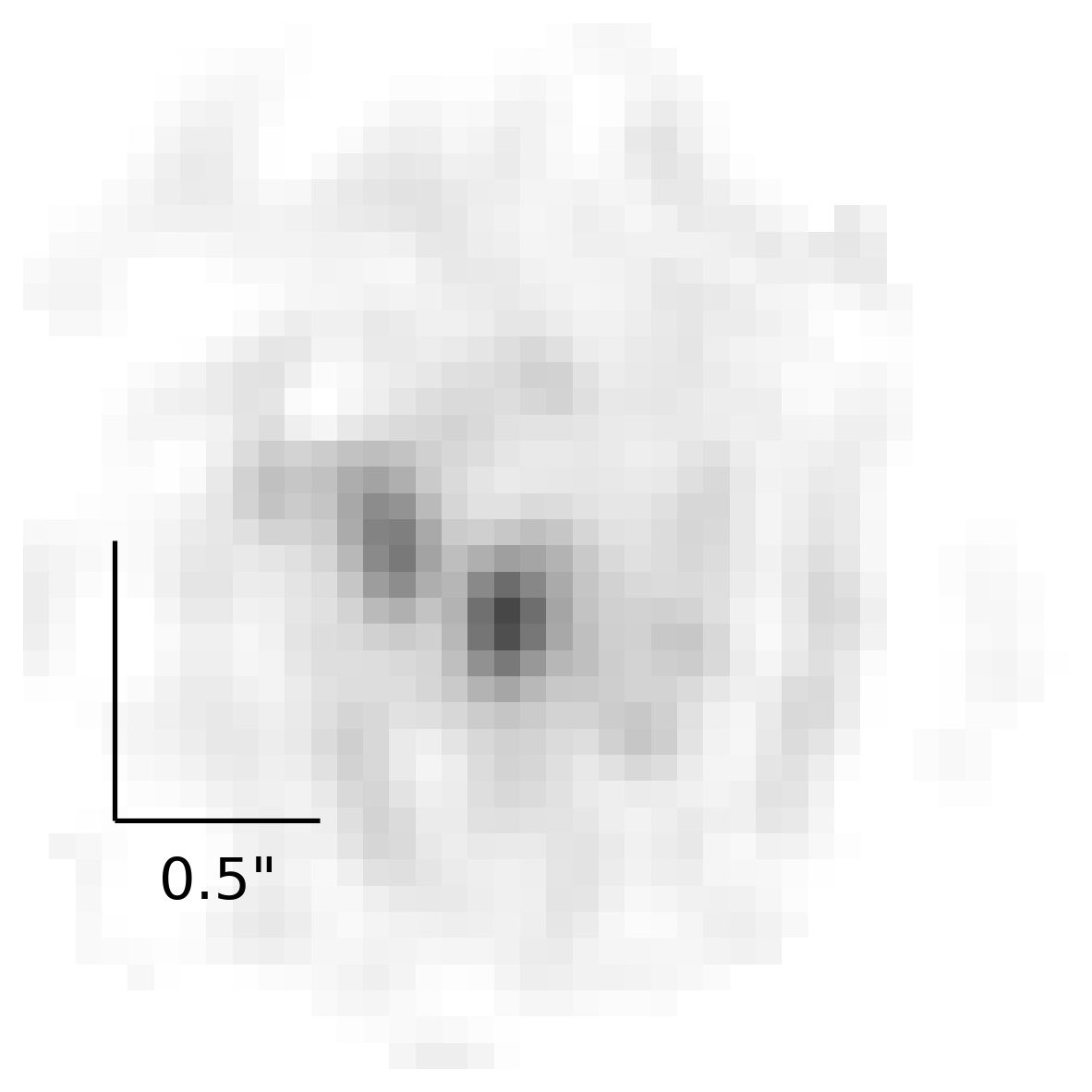}}
   \caption{PIEMD model with one source. From the left to the right: observed SB in the F160W band of the multiple images considered in the extended source modeling; model-predicted SB; normalized residuals in the range from $-5 \sigma$ to $+5 \sigma$; reconstructed SB of the source on a $40 \times 40$ pixel grid. Angular scales of $1 \arcsec$ and of $0.5 \arcsec$ are represented on the lens and source plane, respectively.}
              \label{fig:piemd+sAB40}%
    \end{figure*}

\begin{figure*}
   \centering
   \includegraphics[width=4cm]{figures/ext_piemd_im.jpeg}\hspace{2mm}
   \includegraphics[width=4cm]{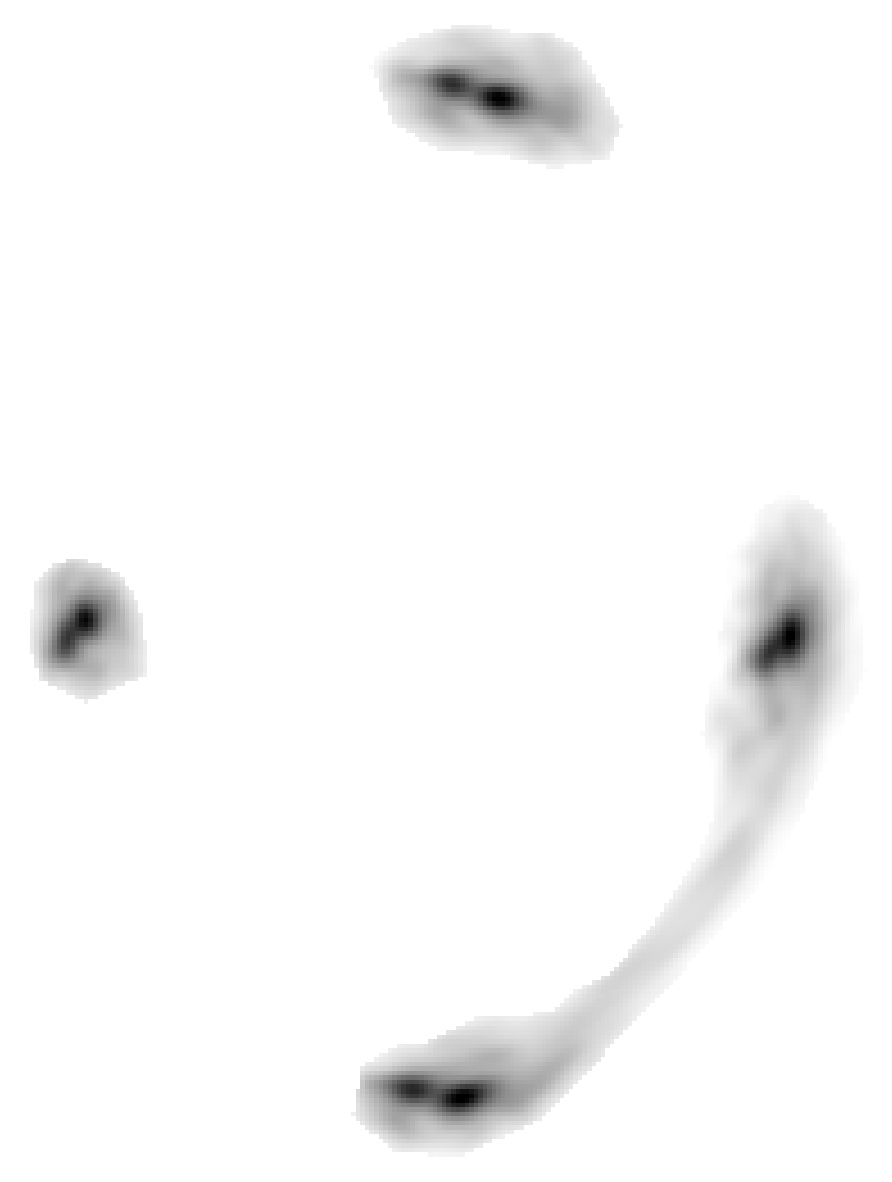}\hspace{2mm}
   \includegraphics[width=5cm]{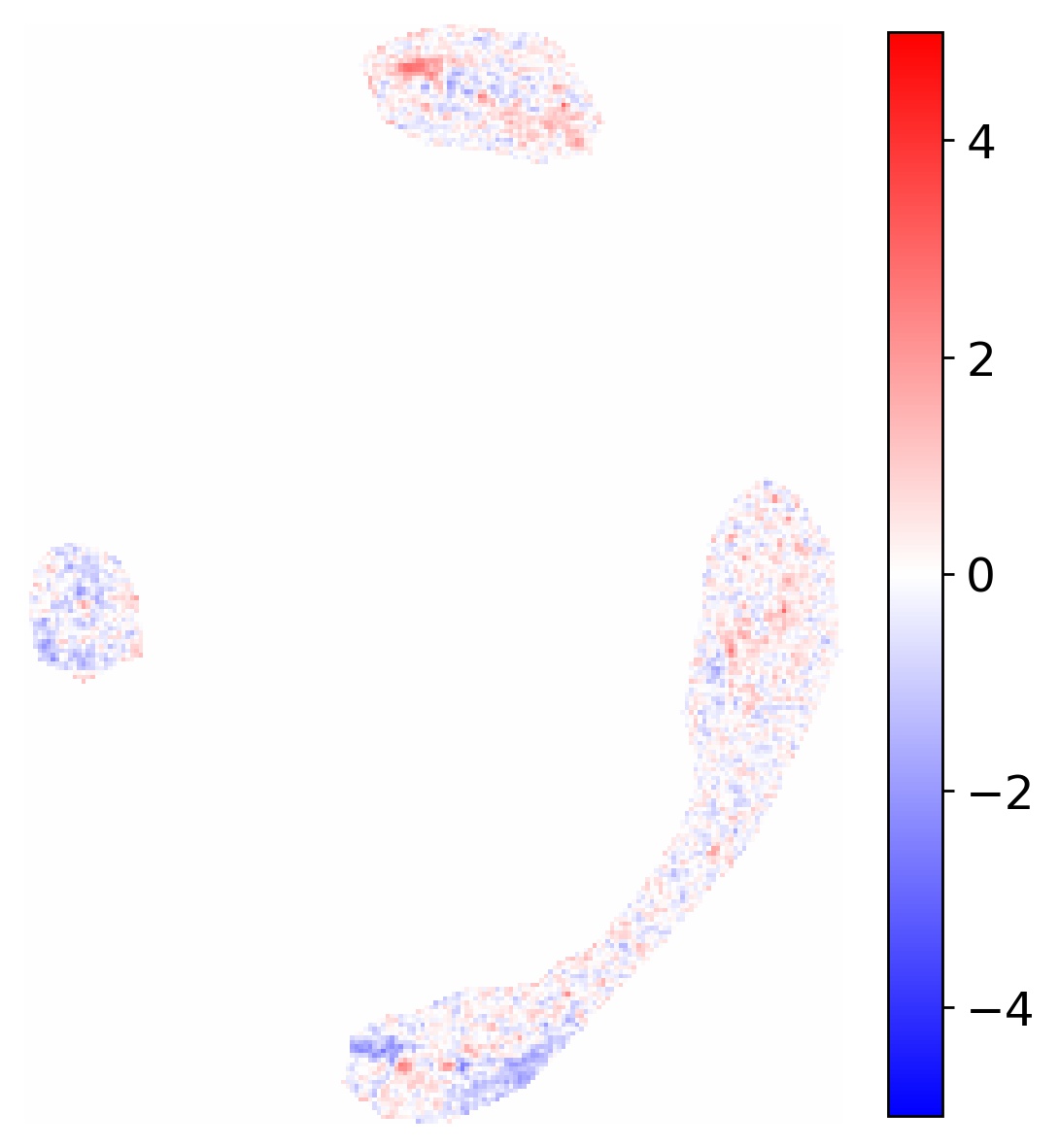}
   \hspace{5mm}
   \raisebox{0.25\height}{\includegraphics[width=3.5cm]{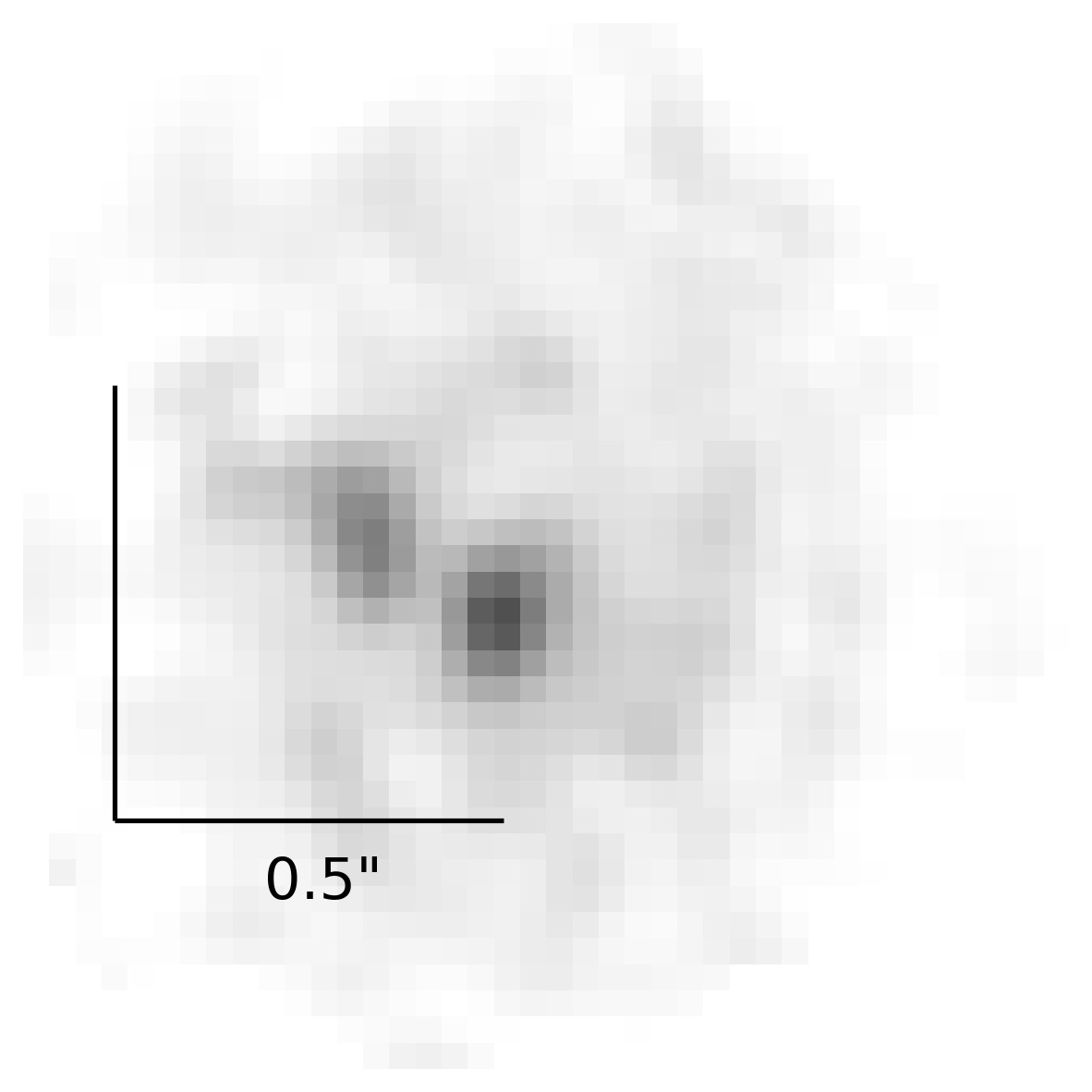}}
   \caption{As in Fig. \ref{fig:piemd+sAB40} for the SPEMD model with one source.}
              \label{fig:spemd+sAB40}%
    \end{figure*}

\begin{figure*}
   \centering
   \includegraphics[width=4cm]{figures/ext_piemd_im.jpeg}\hspace{2mm}
   \includegraphics[width=4cm]{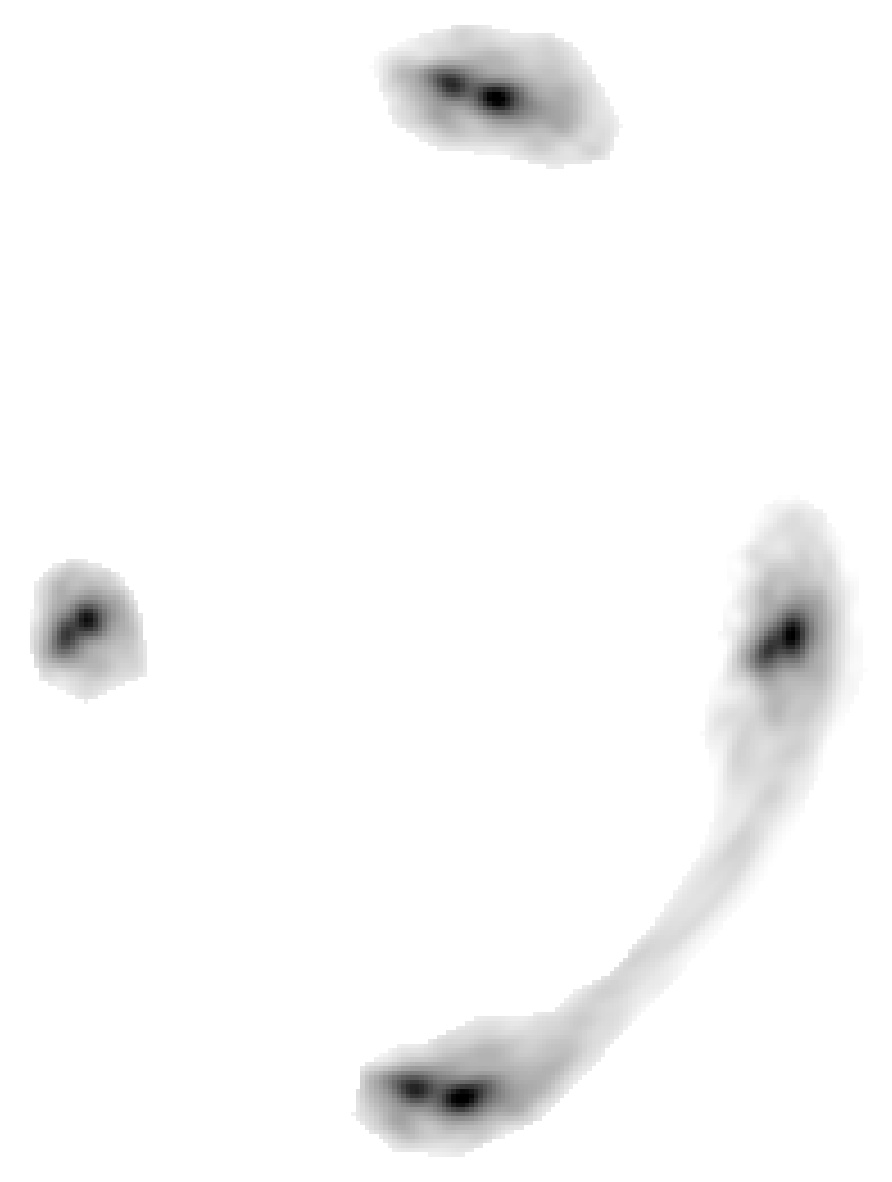}\hspace{2mm}
   \includegraphics[width=5cm]{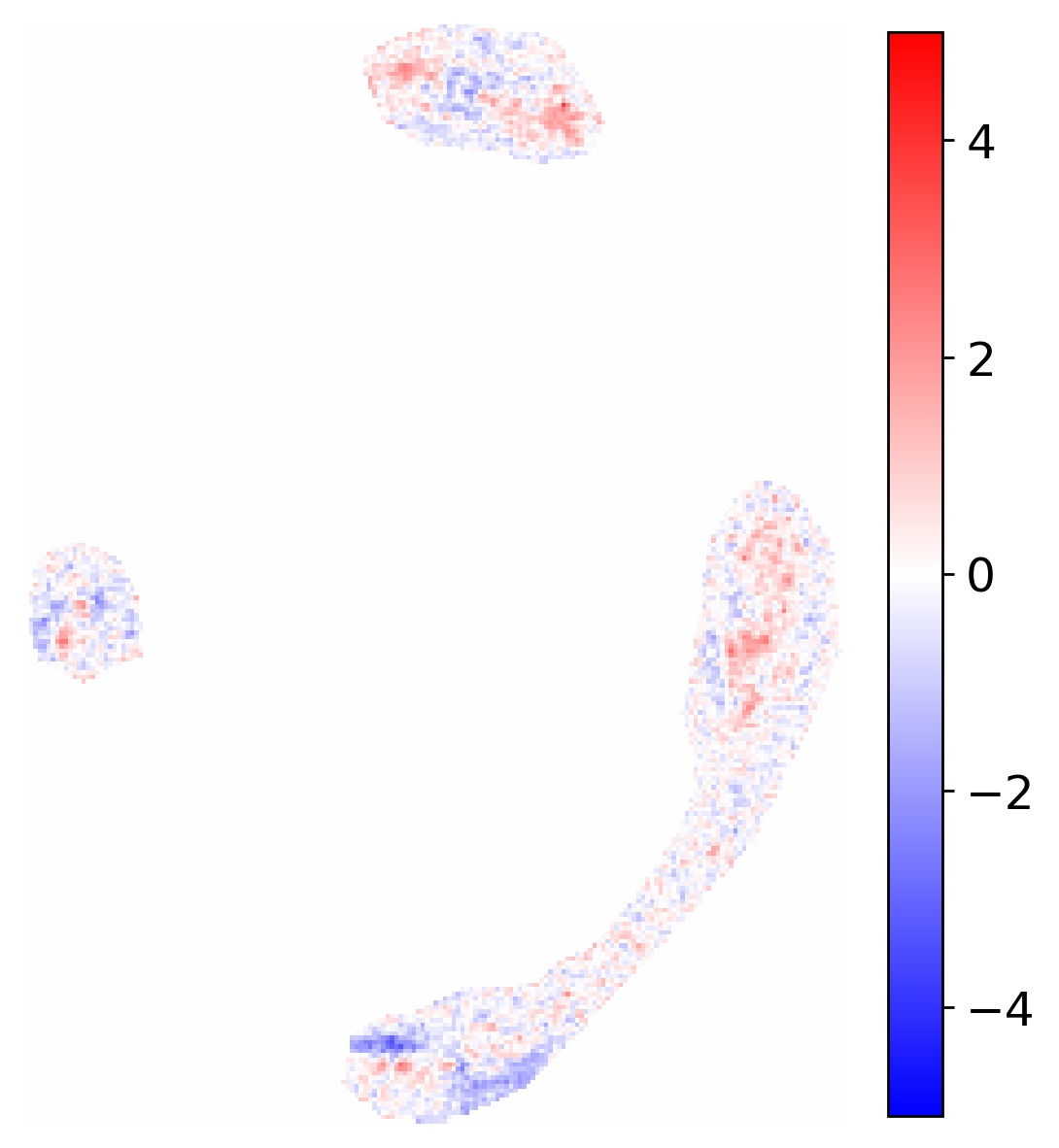}
   \hspace{5mm}
   \raisebox{0.25\height}{\includegraphics[width=3.5cm]{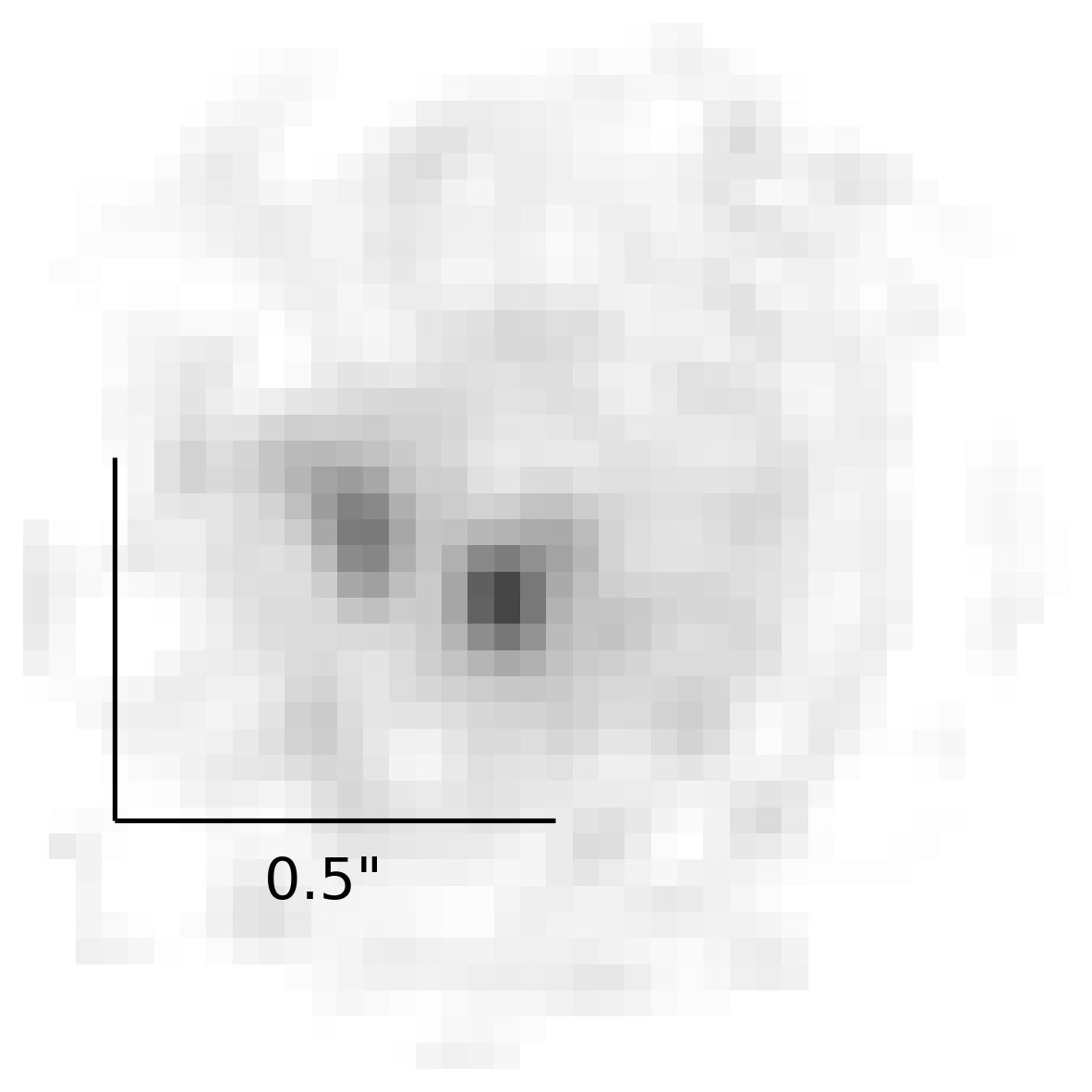}} \\ 
   \includegraphics[width=3.05cm]{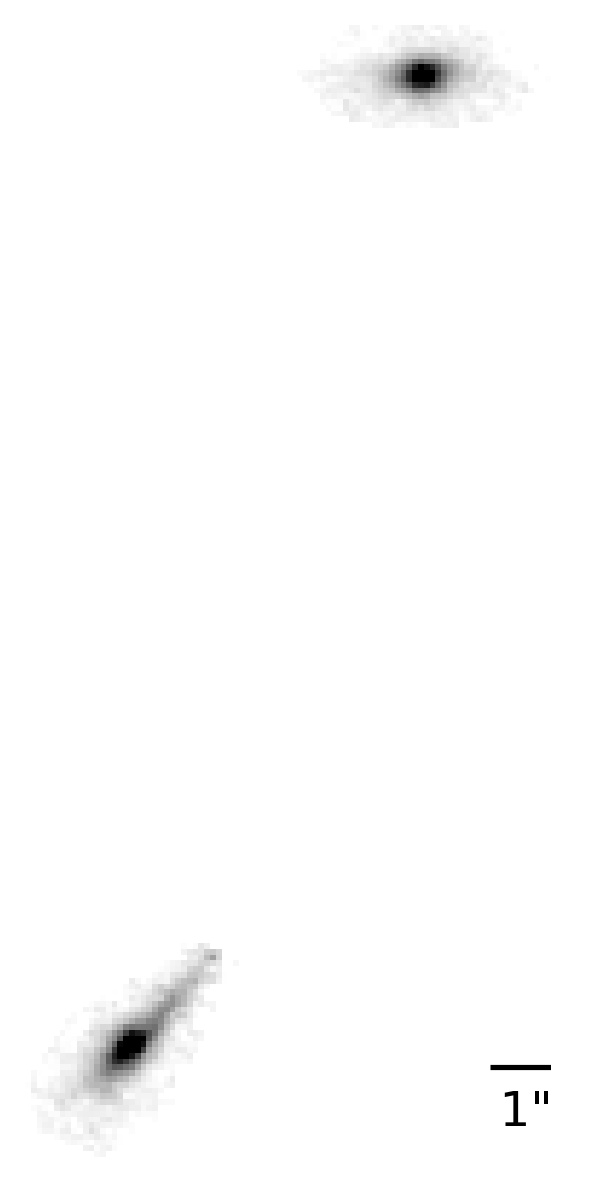}\hspace{19mm}
   \includegraphics[width=2.95cm]{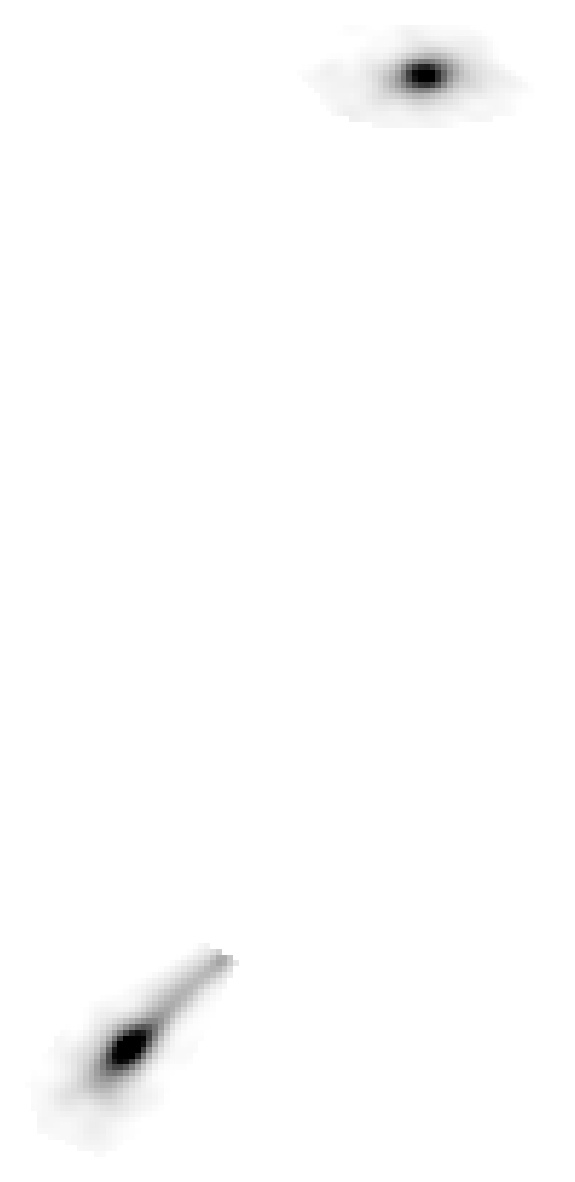}\hspace{19mm}
   \includegraphics[width=3.9cm]{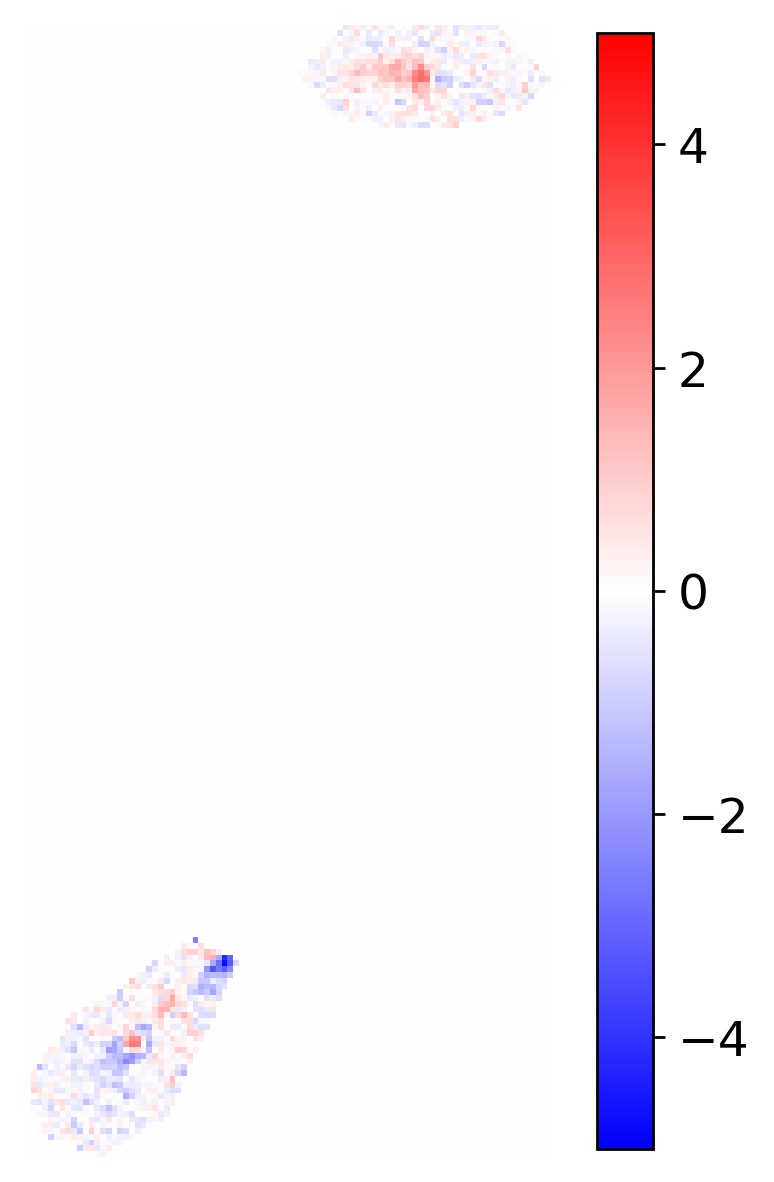}
   \hspace{3mm}
   \raisebox{0.5\height}{\includegraphics[width=3.5cm]{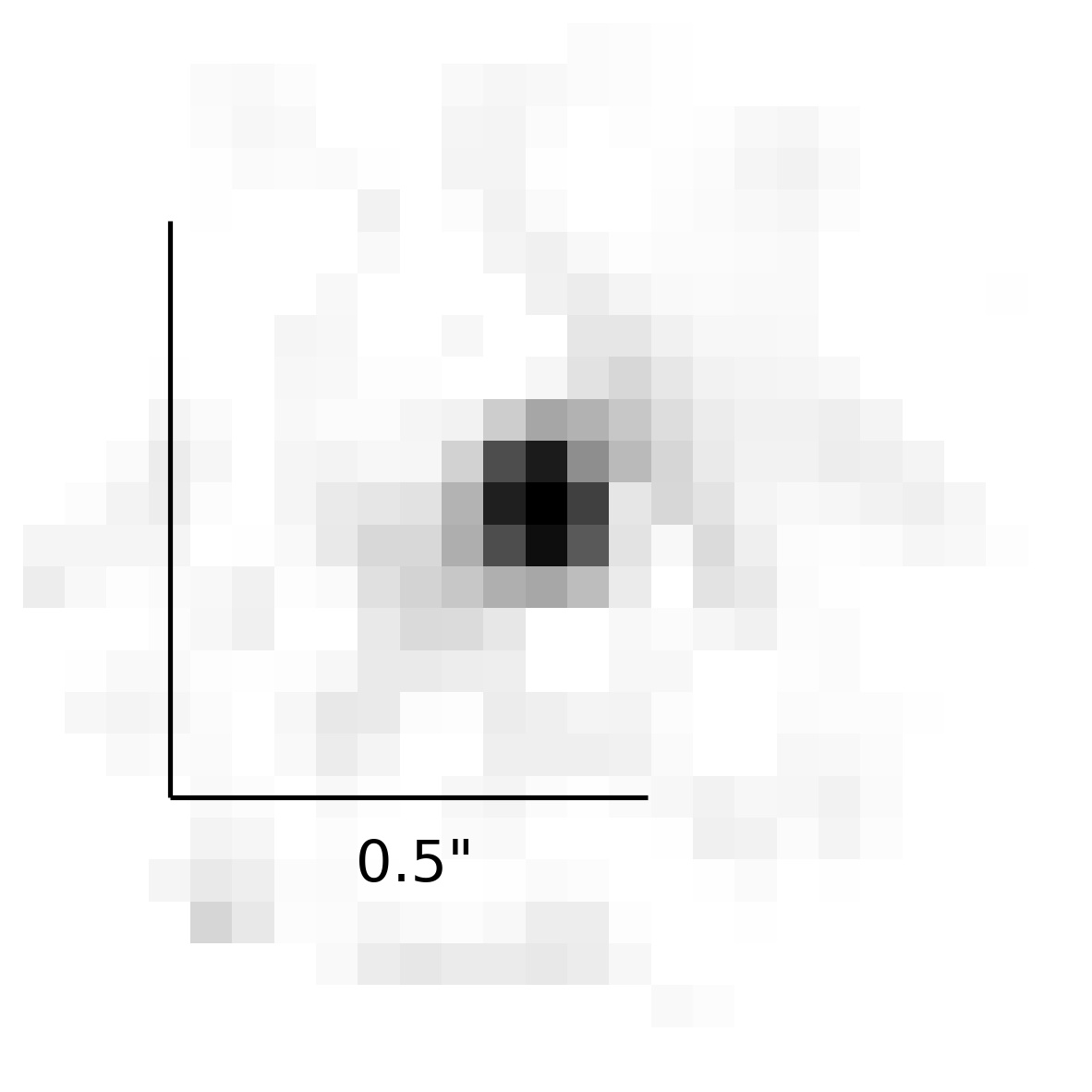}} \\
   \caption{As in Fig. \ref{fig:piemd+sAB40} for the SPEMD model with two sources. The top row shows sources A and B and their SB is reconstructed on a $40 \times 40$ pixel grid, while the bottom row shows source C and its SB is reconstructed on a $25 \times 25$ pixel grid.}
              \label{fig:spemd+sABC}%
    \end{figure*}

Similarly to Figure \ref{fig:comparison}, we show the comparison between the observations and the model predictions for the extended modeling. It is now possible to compare directly, pixel by pixel, the observed and predicted arcs. We show the results for the PIEMD (Fig.~\ref{fig:piemd+sAB40}) and SPEMD (Fig.~\ref{fig:spemd+sAB40}) models when considering the single source AB, and for the SPEMD model when including also source C (Fig.~\ref{fig:spemd+sABC}). In the first panels from the left, we show the pixels of the observed images included in the masks as observables. In the second panels, we plot the extended images predicted by the best-fit models and, in the third panels, the normalized residuals in the ($-5\sigma$, $+5\sigma$) interval. When a single source is reconstructed, we observe that the arcs are very well reproduced with the PIEMD model and the residuals do not exceed in absolute value $ 2.5\sigma$. The results are even better with the SPEMD model, which is able to further reduce the residuals, in particular of the A2B2 and of A4B4 images. Besides the expected increase of the residuals (up to about $3.5\sigma$ around the images A3B3) when the source C is added, the model can predict remarkably well the radial arc (C1). As already discussed, this multiple image is the most sensitive to the deflector inner total mass density profile and its good reconstruction supports the reliability of the strong lensing modeling results and of the lens SB modeling and subtraction. In the panels on the right, we show the reconstructed SB of the sources, which are discussed in Section~\ref{subs:ext_source}. 


\subsection{Total mass profiles of the deflector}   
\begin{figure}[!h]
   \centering
   \includegraphics[width=0.48\textwidth]{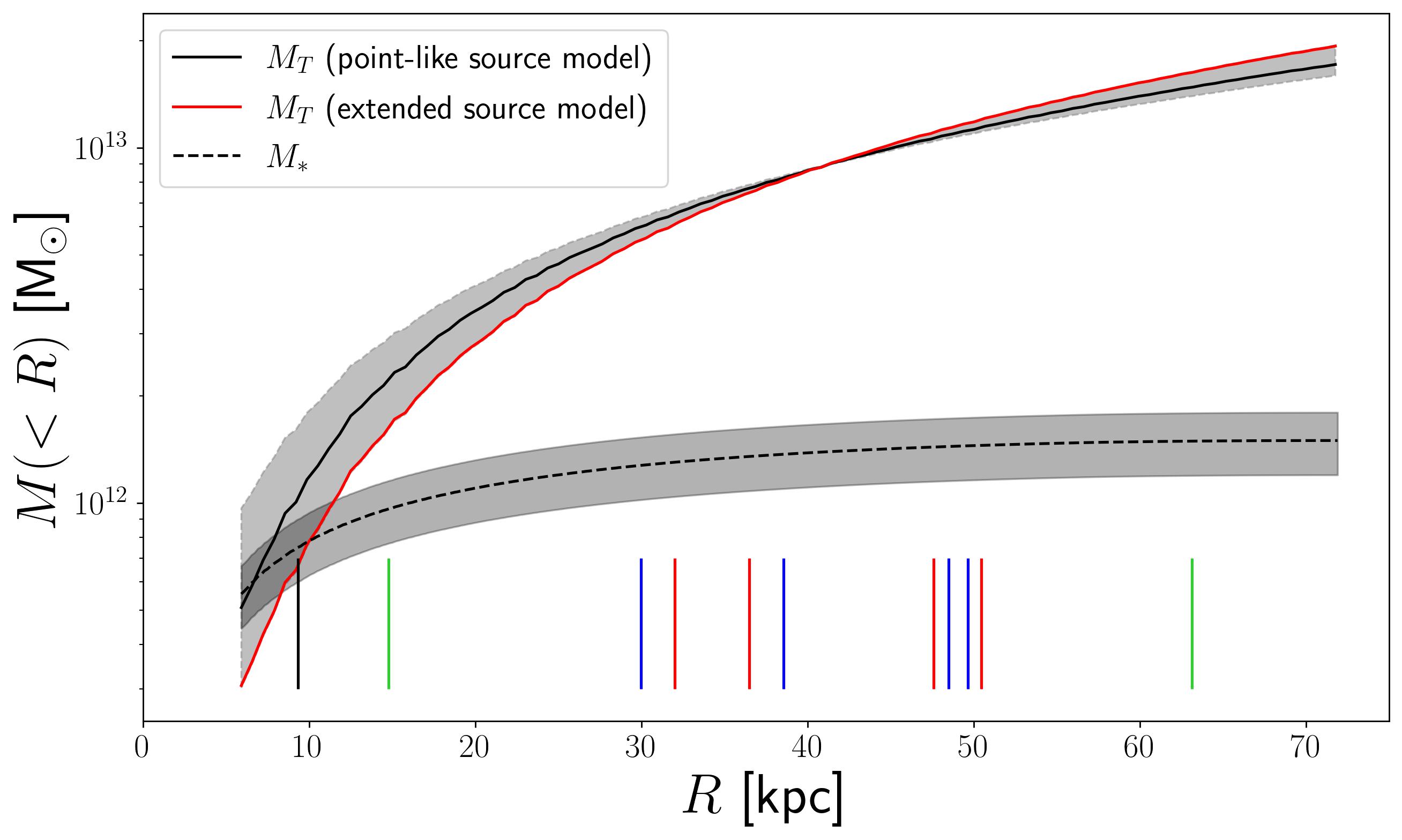}
      \caption{Cumulative projected total mass profiles for the PIEMD$+$rc model with point-like (solid black) and extended source (red) modeling, with $ \pm 1\sigma$ uncertainties (shaded areas), obtained by modeling the multiple images of A, B and C. For the extended source modeling, the uncertainties are smaller than the linewidth. The vertical lines close to the $x$-axis locate the distances from the lens galaxy center of the different multiple images, color-coded following Figs.~\ref{fig:labels} and \ref{fig:comparison}. The black line shows the effective radius of the main lens galaxy. }
         \label{fig:comparison_point-ext}
   \end{figure}

We measure the lens projected total mass distribution from the extended source modeling using the same procedure described for the point-like source modeling. In Fig.~\ref{fig:comparison_point-ext}, we show the cumulative projected total mass profile of the deflector measured with the PIEMD$+$rc model and two extended sources (solid red line, the $1\sigma$ uncertainties are smaller than the linewidth), compared to its corresponding point-like source model (solid black and $1\sigma$ uncertainties as shaded area). The very small statistical uncertainties on the lens total mass profile from the extended source modeling, which are a factor $\gtrsim 10$ smaller than those presented in Sect.~\ref{subs:total_mass_point}, are related to the small uncertainties on the values of all the model parameters discussed above. In addition to the statistical uncertainties, we can estimate the systematic uncertainties by comparing the results of the different models, that fit the data almost equally well. Within $R \approx 42$~kpc, we measure a projected total mass value of $8.76$, $9.00$, $9.08$, and $9.08 \times 10^{12}$~M$_{\odot}$ for the PIEMD and SPEMD (single source), SPEMD and PIEMD+rc (two sources) models, respectively.  

\begin{figure}[!h]
   \centering
   \includegraphics[width=0.48\textwidth]{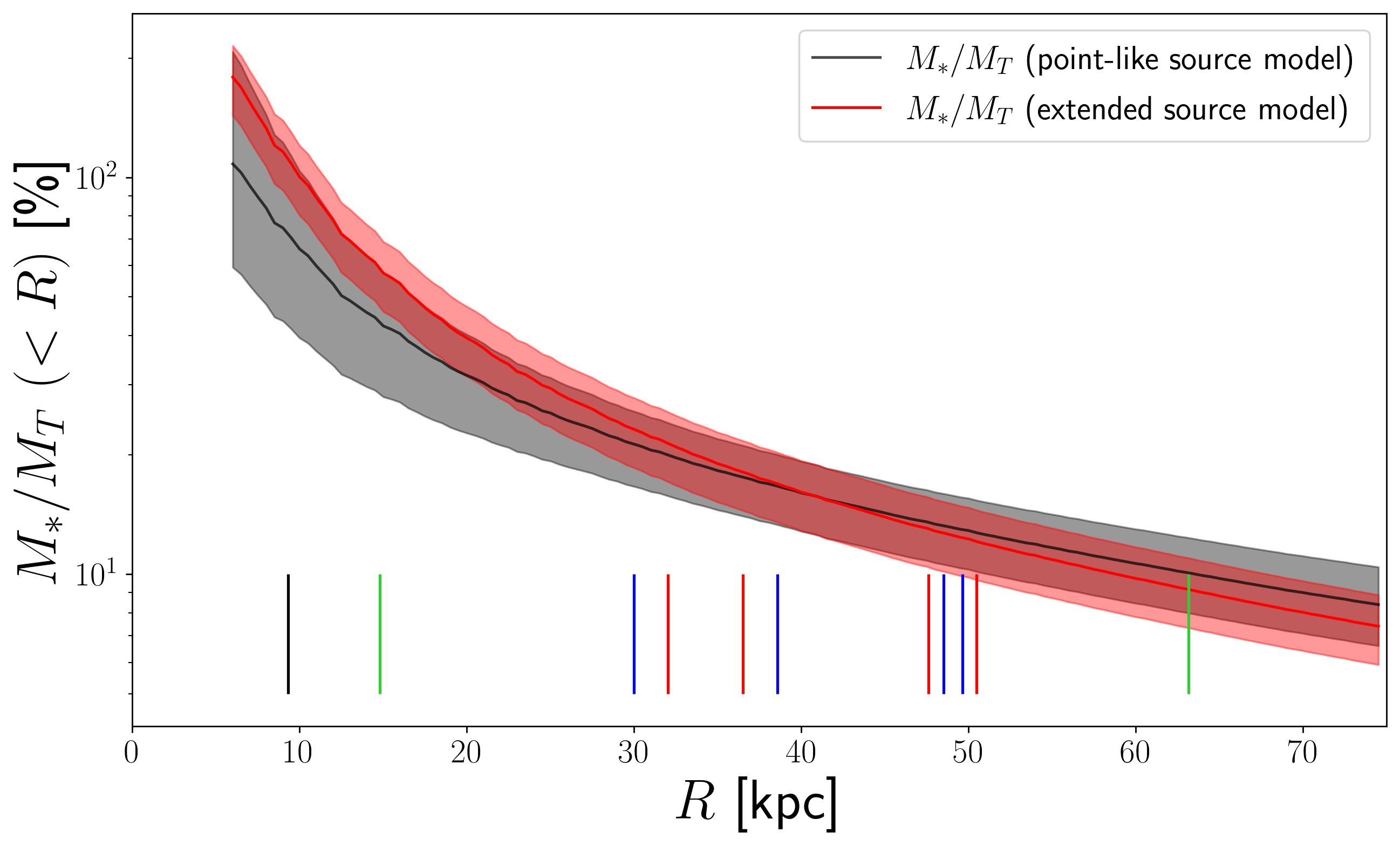}
      \caption{Cumulative projected stellar-over-total mass fraction profiles for the PIEMD$+$rc model with point-like (black) and extended source (red) modeling, with $ \pm 1\sigma$ uncertainties (shaded areas), obtained by modeling the multiple images of A, B and C. The vertical lines are the same as in Fig.~\ref{fig:comparison_piemd-spemd}.}
         \label{fig:ratio_model_ext}
   \end{figure}

On the contrary, the statistical uncertainties on the cumulative projected stellar-over-total mass fraction profiles for corresponding point-like and extended models are very similar, because the uncertainty on the stellar mass profile dominates over that on the total mass profiles. In Fig.~\ref{fig:ratio_model_ext}, we show the profiles for the PIEMD$+$rc models. They present slightly different slopes, but they both result in a stellar-over-total mass fraction value of $(15 \pm 3)\%$ within $42$ kpc. From a comparison with the PIEMD$+$rc point-like model, we find that the extended one predicts a higher value, i.e. $(108 \pm 22)\%$ consistent with one, at the effective radius, and a lower value of $(9 \pm 2)\%$ in the outermost regions ($R \approx 63$~kpc).

\subsection{Source SB reconstruction}
\label{subs:ext_source}
The source SB reconstructions are showed on the fourth panels of Figs.~\ref{fig:piemd+sAB40}, \ref{fig:spemd+sAB40}, and \ref{fig:spemd+sABC}. In each image, there are two different rods along the $x$ and $y$ directions, representing a scale of $0.5\arcsec$, as discussed above. By observing them, we note that, despite recovering the same morphology, the sizes of the background sources vary by a factor of $\sim 2$, when the different total mass profiles for the lens are assumed.

We investigate this difference by measuring the angular separation between the sources A and B, at redshift $z=1.880$, and by estimating the half-light radii of the sources A, B, and C. In the first case, we compare the angular separations predicted by the different point-like and extended models. In particular, the positions of the point-like sources are optimized directly by {\tt GLEE}, while, when considered as extended, we measure the angular separation between the brightest pixels associated to the sources. The results are shown in Tab.~\ref{tab:dimensions sources}. First, we observe that each model predicts approximately the same angular distance, when the point-like or extended modeling is adopted. This test further proves the consistency of the two methods. Then, we confirm quantitatively what we have observed from the reconstructed images: both the PIEMD and SPEMD models can reproduce well the observed multiple images, but they predict angular separations between the sources A and B that differ by a factor of approximately 2. This shows that the choice on the lens total mass distribution has a significant impact on the inferred properties of the background sources. This effect might be very relevant for lens clusters, where hundreds of background lensed sources are detected and the total mass distribution of the lens is more complex. However, in this strong lensing system, even if both the PIEMD and SPEMD models can reproduce successfully the observed multiple images and have approximately the same number of degrees of freedom, the SPEMD model results in a lower $\chi^2_{min}$ value. We remark that the PIEMD$+$rc and SPEMD models, with sources A, B, and C, have the lowest $\chi^2_{min}$ values and predict consistent angular distances. Considering the limited number of different models that can be tested in a strong lensing analysis, we highlight that this effect should be quantified and quoted as a possible source of systematic uncertainty on the reconstructed sources. 

Next, we measure the effective radii of the two components A and B of the first source, and of the source C. We compute the luminosity growth curve of each reconstructed source, measuring the flux included within concentric circular apertures, centered on the brightest pixel, with a step of 0.5 pixels (which corresponds to approximately $0.03 \arcsec$ for all the sources). The apertures are corrected to take into account the different pixel sizes along the $x$ and $y$-axes. For the peaks of the composite AB source, we consider semi-circular apertures, to avoid the contribution of source B to be included in the measurement of the half-light radius of source A, and vice versa. Hence, we build the luminosity cumulative profiles and infer the half-light radii by considering only the halves of the plane less affected by the other peaks. 

We observe that the cumulative luminosity profiles do not converge to a plateau, because of the noise in which the reconstructed sources are immersed. Because of that, we had to choose a maximum radius for the two sources, on the visual inspection of the images, of $4$ and $3$ pixels for the sources A and B, respectively. For source C instead, the cumulative luminosity profile converges to a plateau. The results are summarized on the left part in Tab.~\ref{tab:dimensions sources}. 

We check our results with another test. We analyze the multiple images A1 and B1, which are the less distorted and better resolved, and consider a rectangular cutout of $2.1 \arcsec \times 1.3 \arcsec$ around them. We fit the SB of the background arc with an extended Sérsic component, and the SB of the two images A1 and B1 with two additional Sérsic profiles. From the best-fit model, we measure the values of the effective radii $r_{e\,A1}=0.23\arcsec$ and $r_{e\,B1}=0.17\arcsec$.
Then, we estimate the local magnification factor around the multiple images A1 and B1 to infer their intrinsic sizes on the source plane. For each model, we compute the median magnification map from $1000$ models randomly extracted from the final MCMCs (see Sect.~\ref{subs:total_mass_point}). For each pixel of coordinates ($i,j$) of each model, we compute the local magnification factor as 
\begin{equation}
    \mu_{i,j} = \dfrac{1}{(1-\kappa_{i,j})^2-\gamma_{1\, i,j}^2 - \gamma_{2\, i,j}^2} \, ,
\end{equation}
where $\kappa_{i,j}$, $\gamma_{1\, i,j}$ and $\gamma_{2\, i,j}$ are the values of the convergence and shear components, respectively.
Then, from the 1000 values of magnification estimated on each pixel, we consider the median value and build the median magnification map $\overline{\mu}_{i,j}$. With the 16$^\mathrm{th}$ and 84$^\mathrm{th}$ percentile values of the same distribution, we quantify the $\pm 1\sigma$ uncertainties. Finally, we measure the local magnification factors for the multiple images A1 and B1 by taking the median value within circles centered on them and with radii equal to $r_{e\,A1}$ and $r_{e\,B1}$. The measured local magnification factors around A1 and B1 are reported in Tab.~\ref{tab:dimensions sources}. If we correct the observed effective radii on the lens plane according to
\begin{equation}
\label{eqn:remu}
    r_e^\mathrm{source} = \frac{r_e^\mathrm{image}}{\sqrt{\overline{\mu}}}\, ,
\end{equation}
we find that these results (see the last two columns of Tab.~\ref{tab:dimensions sources}) are consistent, given the uncertainties, with those obtained on the source plane with the luminosity growth curve of the SB reconstructions. 
\begin{table*}
\centering
\small
\begin{tabular}{@{}cc|lcccc|cc|cc@{}}
\toprule
\toprule
 & & Model & $d_{AB} \, [\arcsec] $ & $r_e\, [\arcsec]$  (A) & $r_e\, [\arcsec]$ (B) & \multicolumn{1}{c}{$r_e\, [\arcsec]$ (C)} & $\overline{\mu}_{A1}$ & $\overline{\mu}_{B1}$ & $r_{e, \, A}\, [\arcsec]$ & $r_{e, \, B}\, [\arcsec]$ \\
\midrule
\parbox[t]{2mm}{\multirow{8}{*}{\rotatebox[origin=c]{90}{Point-like}}}  & \parbox[t]{2mm}{\multirow{4}{*}{\rotatebox[origin=c]{90}{A,B}}} & {\scriptsize PIEMD} & 0.34  & & & & $2.59^{+0.12}_{-0.10}$ & $2.49^{+0.11}_{-0.09}$ & $0.14^{+0.01}_{-0.01}$  &  $0.11^{+0.01}_{-0.01}$ \\
& & & (2.87 kpc) & & & & & & ($1.18^{+0.02}_{-0.03}$ kpc) & ($0.89^{+0.02}_{-0.02}$ kpc) \\
& & {\scriptsize SPEMD} & 0.19  & & & & $3.84^{+4.46}_{-1.21}$  & $3.68^{+4.13}_{-1.14}$ & $0.12^{+0.02}_{-0.07}$  &  $0.09^{+0.01}_{-0.05}$  \\
&&& (1.58 kpc) & & & & & & ($0.97^{+0.15}_{-0.56}$ kpc) & ($0.73^{+0.11}_{-0.41}$ kpc) \\
\cmidrule{2-11} 
& \parbox[t]{1.5mm}{\multirow{4}{*}{\rotatebox[origin=c]{90}{A,B,C}}}  & {\scriptsize PIEMD+rc} & 0.19  & & & & $3.60^{+4.44}_{-1.15}$ & $3.44^{+4.06}_{-1.07}$ & $0.12^{+0.02}_{-0.07}$  &  $0.09^{+0.01}_{-0.05}$  \\
&&& (1.62 kpc) & & & & & & ($1.00^{+0.16}_{-0.62}$ kpc) & ($0.75^{+0.12}_{-0.44}$ kpc) \\
& & {\scriptsize SPEMD} & 0.19  & & & & $3.21^{+0.79}_{-0.58}$ &$3.11^{+0.75}_{-0.55}$ & $0.13^{+0.01}_{-0.02}$  &  $0.09^{+0.01}_{-0.01}$  \\
&&& (1.61 kpc) & & & & & & ($1.06^{+0.10}_{-0.13}$ kpc) & ($0.79^{+0.07}_{-0.10}$ kpc) \\
\midrule \midrule
\parbox[t]{2mm}{\multirow{8}{*}{\rotatebox[origin=c]{90}{Extended}}} & \parbox[t]{2mm}{\multirow{4}{*}{\rotatebox[origin=c]{90}{A,B}}}  & {\scriptsize PIEMD} & 0.28 & 0.14  & 0.10  & & $2.80^{+0.10}_{-0.10}$ & $2.69^{+0.01}_{-0.01}$ & $0.14^{+0.01}_{-0.01}$  &  $0.10^{+0.01}_{-0.01}$   \\
&&& (2.41 kpc) & (1.18 kpc) & (0.87 kpc) &&&& ($1.14^{+0.02}_{-0.02}$ kpc) & ($0.85^{+0.01}_{-0.01}$ kpc) \\
& & {\scriptsize SPEMD} & 0.18  & 0.08  & 0.06   & &  $7.27^{+0.50}_{-0.28}$ & $6.93^{+0.47}_{-0.26}$ & $0.08^{+0.01}_{-0.01}$  &  $0.06^{+0.01}_{-0.01}$  \\
&&& (1.55 kpc) & (0.70 kpc) & (0.51 kpc) && && ($0.71^{+0.01}_{-0.02}$ kpc) & ($0.53^{+0.01}_{-0.02}$ kpc) \\
\cmidrule{2-11} 
& \parbox[t]{1.5mm}{\multirow{4}{*}{\rotatebox[origin=c]{90}{A,B,C}}}
& {\scriptsize PIEMD+rc} & 0.16  & 0.07  & 0.05  & 0.10  & $7.64^{+0.08}_{-0.09}$ & $7.16^{+0.07}_{-0.12}$ & 
$0.08^{+0.01}_{-0.01}$ &  $0.06^{+0.01}_{-0.01}$ \\
&&& (1.36 kpc) & (0.60 kpc) & (0.46 kpc) & (0.81 kpc, $z=1.69$) & & & ($0.69^{+0.01}_{-0.01}$ kpc) &  ($0.52^{+0.01}_{-0.01}$ kpc) \\
& & {\scriptsize SPEMD} & 0.18  & 0.09  & 0.06  & 0.06  & $6.22^{+0.08}_{-0.06}$ & $5.96^{+0.07}_{-0.06}$ &
$0.09^{+0.01}_{-0.01}$  &  $0.07^{+0.01}_{-0.01}$ \\
&&& (1.49 kpc) & (0.75 kpc) & (0.52 kpc) & (0.50 kpc, $z=2.04$) &&& ($0.76^{+0.01}_{-0.01}$ kpc) & ($0.57^{+0.01}_{-0.01}$ kpc) \\
\bottomrule
\end{tabular}%
\caption{First column: angular separation between the sources A and B. In the point-like source modeling (top), this is computed as the distance between the best-fit source positions optimized by {\tt GLEE}, while in the extended modeling (bottom), it is computed as the distance between the brightest pixels associated to each source in the reconstructed SBs. The second, third and fourth columns show the half-light radii of the reconstructed sources estimated through their cumulative luminosity profiles. Fifth and sixth columns: local median magnification factors of the multiple images A1 and B1 and $1\sigma$ statistical uncertainties. They are measured as the median value of the median magnification maps within circles of radii $r_{e\,A1}$ and $r_{e\,B1}$, centered on A1 and B1. Last two columns: half-light radii of the sources A and B measured from the results $r_{e\,A1}=0.23\arcsec$ and $r_{e\,B1}=0.17\arcsec$ of the SB modeling of the multiple images A1 and B1, and demagnified with Eq. (\ref{eqn:remu}). The uncertainties are propagated from those on the magnification factors. The angular quantities are converted to physical ones for $z=1.880$, unless otherwise specified.  }            
\label{tab:dimensions sources}  
\end{table*}
\begin{table*}
\centering
\small
\begin{tabular}{@{}cc|llccccccccc@{}}
\toprule
\toprule
 & & Parameters & $x$ {[}$''${]} & $y$ {[}$''${]} & $q$ & $\theta$ {[}rad{]} & $\theta_E$ {[}$''${]} & $r_{core}$ {[}$''${]} & $\gamma'$ & $\gamma_\mathrm{ext}$ & $\phi_\mathrm{ext}$ {[}rad{]} & $z_C$\\
\midrule
\parbox[t]{2mm}{\multirow{4}{*}{\rotatebox[origin=c]{90}{Point-like}}}  & \parbox[t]{2mm}{\multirow{2}{*}{\rotatebox[origin=c]{90}{A,B}}} & PIEMD & $0.27_{-0.03}^{+0.03}$ & $-0.14_{-0.03}^{+0.02}$ & $0.67_{-0.10}^{+0.09}$ & $0.07_{-0.03}^{+0.05}$ & $11.2_{-0.2}^{+0.4}$ & $[0.0]$ & & $0.20_{-0.05}^{+0.04}$ & $1.44_{-0.03}^{+0.02}$ & \\
& & SPEMD & $0.26_{-0.07}^{+0.06}$ & $-0.15_{-0.06}^{+0.05}$ & $0.61_{-0.16}^{+0.18}$ & $0.05_{-0.04}^{+0.09}$ & $7.0_{-2.0}^{+2.9}$ & $[0.0]$ & $2.01_{-0.24}^{+0.17}$ & $0.18_{-0.09}^{+0.09}$ & $1.43_{-0.15}^{+0.05}$ &\\
\cmidrule{2-13} 
& \parbox[t]{1.5mm}{\multirow{2}{*}{\rotatebox[origin=c]{90}{\scriptsize{A,B,C}}}}  & PIEMD+rc & $0.18_{-0.05}^{+0.05}$ & $-0.12_{-0.06}^{+0.05}$ & $0.74_{-0.21}^{+0.12}$ & $0.07_{-0.05}^{+0.10}$ & $15.6_{-2.6}^{+3.9}$ & $2.1_{-1.6}^{+1.6}$ & &$0.09_{-0.03}^{+0.03}$ & $1.31_{-0.12}^{+0.08}$ &$1.88_{-0.19}^{+0.38}$ \\
& & SPEMD & $0.16_{-0.03}^{+0.03}$ & $-0.19_{-0.03}^{+0.03}$ & $0.48_{-0.07}^{+0.07}$ & $0.02_{-0.01}^{+0.01}$ & $6.8_{-1.3}^{+1.6}$ & [0.0] & $1.85_{-0.11}^{+0.10}$ & $0.05_{-0.01}^{+0.01}$ & $0.70_{-0.29}^{+0.37}$ & $1.48_{-0.08}^{+0.09}$ \\
\midrule \midrule
\parbox[t]{2mm}{\multirow{4}{*}{\rotatebox[origin=c]{90}{Extended}}} & \parbox[t]{2mm}{\multirow{2}{*}{\rotatebox[origin=c]{90}{A,B}}}  & PIEMD & $0.24_{-0.01}^{+0.01}$ & $-0.14_{-0.01}^{+0.01}$ & $0.79_{-0.01}^{+0.01}$ & $0.14_{-0.01}^{+0.01}$ & $10.9_{-0.1}^{+0.1}$ & $[0.0]$ & & $0.25_{-0.01}^{+0.01}$ & $1.45_{-0.01}^{+0.01}$ & \\
& & SPEMD & $0.18_{-0.01}^{+0.01}$ & $-0.14_{-0.01}^{+0.01}$ & $0.76_{-0.01}^{+0.01}$ & $0.09_{-0.01}^{+0.01}$ & $3.9_{-0.1}^{+0.1}$ & $[0.0]$ & $1.57_{-0.02}^{+0.02}$ & $0.06_{-0.01}^{+0.01}$ & $1.06_{-0.08}^{+0.06}$ & \\
\cmidrule{2-13} 
& \parbox[t]{1.5mm}{\multirow{2}{*}{\rotatebox[origin=c]{90}{\scriptsize{A,B,C}}}}
& PIEMD+rc & $0.09_{-0.01}^{+0.01}$ & $-0.13_{-0.01}^{+0.01}$ & $0.84_{-0.01}^{+0.01}$ & $0.15_{-0.01}^{+0.01}$ & $18.7_{-0.1}^{+0.1}$ & $3.4_{-0.1}^{+0.1}$ & & $0.09_{-0.01}^{+0.01}$ & $1.25_{-0.01}^{+0.01}$ & $2.04_{-0.01}^{+0.02}$ \\
& & SPEMD & $0.05_{-0.01}^{+0.01}$ & $-0.17_{-0.01}^{+0.01}$ & $0.62_{-0.01}^{+0.01}$ & $0.04_{-0.01}^{+0.01}$ & $4.2_{-0.1}^{+0.1}$ & [0.0] &$1.58_{-0.01}^{+0.01}$ & $0.08_{-0.01}^{+0.01}$ & $0.28_{-0.01}^{+0.01}$ & $1.69_{-0.01}^{+0.01}$ \\
\bottomrule
\end{tabular}%
\caption{Median and $68\%$ confidence level uncertainty values of the parameters of the different models, based on the point-like (top) and extended (bottom) modeling of the sources, as described in Tabs.~\ref{tab:point models} and \ref{tab:extended model}. Statistics are extracted from the final MCMC chains with $10^6$ steps. To compare properly different models with different numbers of observables and degrees of freedom, we have rescaled the errors on the observed multiple images for each model, so that its $\chi^2_{min}$ value is approximately equal to that of the number of $\mathrm{dof}$.}             
\label{tab:mcmc}  
\end{table*}

\section{Discussion and summary}
\label{sec:discussion}
In this paper, we have studied SDSS\,J0100+1818, a strong lensing system consisting of a massive early-type galaxy, with a spectroscopic redshift of $z=0.581$, which acts as a gravitational lens on two background sources, AB and C. One of the sources (AB), spectroscopically confirmed at $z=1.880$, has four multiple images, visible around the deflector at a projected distance of about $7 \arcsec$, and presents two components.
The other source (C), instead, does not have a spectroscopic redshift measurement, and its two multiple images represent the closest and the most distant images from the deflector center. Thus, the introduction of this source is key to the reconstruction of the total mass profile of the lens in its inner and outer regions, approximately from $15$ to $63$ kpc, and to the reduction of some degeneracies among the model parameters.

We have developed several strong lensing models of the deflector with the software {\tt GLEE}, combining (cored or singular) PIEMD and SPEMD total mass profiles with or without an external shear component. At the beginning, we have considered the sources as point-like objects, and modeled the lensed source positions. We took advantage of the eight observed positions of the source AB (four multiple images for each of the two components) and of the two observed positions of the source C. The redshift of source C has also been included as a free parameter in the models. 

Then, we have considered the sources as extended objects, and reconstructed their SB distributions. With this improvement, we have been able to exploit the image structure and the extended arcs in which they are distorted, over $\sim$7200 \textit{HST} pixels. We have finally used the reconstructed sources to measure their sizes and discussed how much they can be affected by the adopted total mass profile for the deflector. 

The main results of this work can be summarized as follows: 
\begin{itemize}
    \item We have combined the available multiband photometry from PanSTARRS, NOT, and \textit{HST} to model the spectral energy distribution of the main lens galaxy. The best-fit model results in a stellar mass value of $(1.5 \pm 0.3) \times 10^{12}$~M$_{\odot}$. By using the public software GALFIT on the \textit{HST} image in the F160W band of the system, we have modeled the light distribution of the lens with a combination of two Sérsic profiles. Starting from them, we have measured the cumulative luminosity profile, then converted into a stellar mass profile, by assuming a constant stellar mass-to-light-ratio. \\
    
    \item We have used the public software pPXF to estimate the value of the stellar velocity dispersion $\sigma$ of the main lens galaxy from its X-Shooter spectrum. We have measured that $\sigma = (451 \pm 37)$ km~s$^{-1}$, which is consistent with the very large values of the galaxy stellar mass and of the mean distance between the observed multiple images, confirming that SDSS\,J0100+1818 is among the rarest, most massive elliptical galaxies \citep{Loeb2003}. \\

    \item With the point-like source modeling, we have found a total mass value projected within the Einstein radius (of approximately 42~kpc) of $(9.1 \pm 0.1) \times 10^{12}$~M$_\odot$, consistent for the PIEMD$+$rc and SPEMD models (Fig.~\ref{fig:comparison_piemd-spemd}). The source C is predicted at $z_C=1.72$ ($1.98$) for the PIEMD$+$rc (SPEMD) model (Tab.~\ref{tab:point models}). The best-fit value of the logarithmic slope of the SPEMD model is shallower than for a singular isothermal profile. However, we remark the observation of the expected degeneracies between the values of the $\gamma^\prime$, $\theta_E$, and $z_C$ parameters (Fig.~\ref{fig:corner_models}). \\
    
    \item With the extended source modeling, we have confirmed a projected total mass value enclosed within the Einstein radius of $9.1 \times 10^{12}$~M$_\odot$ both with the PIEMD$+$rc and SPEMD models. The source C is predicted at $z_C=1.69$ ($2.04$) for the PIEMD$+$rc (SPEMD) model (Tab.~\ref{tab:extended model}). By considering the extended structure of the sources and of their multiple images, the number of observables increases to approximately 7200 (against the 20 observables of the point-like source modeling). As a consequence, the statistical uncertainties on the parameters values and on the derived quantities are strongly reduced (see Fig.~\ref{fig:comparison_point-ext}). 
    
    
    \item In \cite{Dutton2014} and \cite{Newman2015}, the value of $\gamma_\mathrm{tot}$ was defined as the three-dimensional mass-weighted mean value of the density slope within $R_e$ and it was discussed how this varies by considering a sample of 59 galaxy-scale lenses from the SLACS survey ($\langle z \rangle~=~0.20$; \citealt{Auger2009, Auger2010a}), 10 group-scale lenses ($\langle z \rangle~=~0.36$), and 7 central galaxies of massive clusters (BCGs, $\langle z \rangle~=~0.25$; \citealt{Newman2013a, Newman2013b}). We have measured the value of $\gamma_\mathrm{tot}$ from the projected total mass profiles $M(R)$ presented in Sect.~\ref{sec:point} and \ref{sec:extended}. Thus, we can directly compare the results only for the models with a constant slope, i.e., for a SPEMD profile with $r_{core}=0$. We have obtained a $\gamma_\mathrm{tot}$ value of $1.89^{+0.22}_{-0.09}$ for the point-like source SPEMD model and of $1.68^{+0.01}_{-0.01}$ for the extended source SPEMD model, considering the A, B, C background sources. In both cases, we have observed that the introduction of the source C slightly increases the value of $\gamma_\mathrm{tot}$, by 0.03 and 0.05, respectively. 
    These values are consistent with those found by \cite{Newman2015} for group-scale systems with $R_e~\approx~10$ kpc, which lie between the values obtained for galaxy-scale ($2.09 \pm 0.03$) and cluster-scale ($1.18 \pm 0.07^{+0.05}_{-0.07}$) systems. \\
    
    \item \cite{Newman2015} exploited the samples introduced above to compare the values of the projected stellar over total mass fraction within $R_e$, defined as
    \begin{equation}
        f_{*, \, \mathrm{Salp}} = \frac{M_*(R_e)}{M_T(R_e)} \, ,
    \end{equation}
    where the subscript clarifies that the stellar mass values were estimated assuming a Salpeter stellar IMF. They showed that also the value of the projected stellar over total mass fraction decreases with increasing Einstein radius and halo mass, and that galaxy, group and cluster-scale systems populate different regions in these parameter spaces. They measured a mean projected stellar over total fraction value of $0.60$ for galaxy-scale lenses, $0.17$ for group-scale lenses, and $0.06$ for cluster-scale lenses. In this work, for SDSS\,J0100+1818, we measured fractions of $0.70 \pm 0.29$ and $0.49 \pm 0.12$ for the PIEMD+rc and SPEMD models, respectively (see Fig.~\ref{fig:ratio_model}). \\ 
    
    \item We have explored the possible presence of other group members and estimated photometric redshifts over the 2.5\arcmin\,$\times$\,2.5\arcmin\, field-of-view covered by our multiband images. It has not been possible to gather quantitative conclusions, due to the lacking spectroscopy. However, we have qualitatively observed an overdensity of candidate members in the north-east direction, which corresponds to a shear position angle $\phi_\mathrm{ext}$ of about $20 \degree$ (note the different orientation of the compass in Fig.~\ref{fig:labels} and of the {\tt GLEE} $x$-axis). This is in relatively good agreement with the best-fit values of $\phi_\mathrm{ext}$, which range from $16 \degree$ to $71 \degree$. We note that the value of $\phi_\mathrm{ext}$ shows some expected degeneracies, mainly with the values of $b/a$, $\theta$, and $\gamma_\mathrm{ext}$.    \\
    
    \item We have reconstructed the SB distributions of the background sources and measured their half-light radii from the luminosity profiles. We have successfully recovered the two-peaked structure of the AB source, with a small physical separation between 1.4 and 2.9 kpc (at $z=1.880$) from all the models (Tab.~\ref{tab:dimensions sources}). We have found that considering different models with similar $\chi^2_{min}/\mathrm{dof}$ values the sizes of the reconstructed sources can vary by a factor of about $2$. These results can have important consequences on the strong lensing modeling of more complicated lens mass distributions, i.e. on cluster scales. However, we remark that in this study the models with a SPEMD or a PIEMD$+$rc lens total mass profile have the lowest $\chi^2_{min}$ values and they predict reconstructed sources with consistent sizes (confirmed also with the extended source modeling). \\

    \item We have measured values of the effective radius between $0.5$ and $1$ kpc at $z~=~1.880$ for the A and B components, depending on the adopted model. Approximately 60\% of $z\sim 2$ galaxies show bright star-forming regions, dubbed \textit{clumps} \citep[e.g.,][]{Cowie1995, Elmegreen2005, Elmegreen2009, 2013ApJ...774...86E, Guo2015, Guo2018, Zanella2019}. Our measured sizes are in agreement with those of strongly lensed clumps in star-forming galaxies at $z\sim~1-3$, with similar moderately-magnified sources $(\mu \sim 3-10)$. Hydrodynamic simulations suggest that the measured clump sizes depend on the spatial resolution of the observations \citep[e.g.,][]{Oklopic2017, Behrendt2016, Behrendt2019, Tamburello2017, Faure2021}, and thus that the measured size upper limit decreases with an increasing magnification factor \citep{Mestric2022, Claeyssens2022}. In high amplification regimes, recent observations have been able to explore clumps with sizes of $\sim 100$ pc \citep{Dessauges-Zavadsky2019, Livermore2015, Cava2018}, down to $\lesssim~20$ pc in some extremely magnified cases \citep{Rigby2017, Vanzella2020, Mestric2022, Claeyssens2022, Messa2022}. Hence, considering the measured magnification factors for the multiple images of the SDSS\,J0100+1818 system, it remains unclear whether A and B are monolithic, isolated clumps, or blends of smaller clumps/sub-components unresolved with \textit{HST}.
    \\

\end{itemize}

   This study could be improved in several ways with additional integral field spectroscopy. With these data, we could: 1)~measure the redshift value of the source C, that would be crucial to break the degeneracy between the values of $\theta_E$ and $\gamma^\prime$. A measurement of the lens total mass enclosed within two (different) Einstein radii would indeed allow us to estimate the values of $\theta_E$ and $\gamma^\prime$ with a precision of a few percent. The measured redshift of C would also allow us to perform a multi-plane lensing analysis, in which the light emitted by the source AB is also deflected by the total mass distribution of C (or viceversa, if $z_C>1.88$); 2)~confirm or reject the hypothesis of the group nature of SDSS\,J0100+1818. If some neighbor galaxies were confirmed at redshifts similar to that of the main deflector, we would be able to include them individually in the strong lensing model (and not simply as an external shear component); 3) measure the velocity dispersion profile of the main lens galaxy and combine kinematics and strong lensing information. The previous points would also pave the way to the confirmation and inclusion of the additional background source we have mentioned in Sec.~\ref{sec:system}. If confirmed, the introduction of another source, at a different redshift, strongly lensed into four multiple images, would importantly improve our strong lensing model and make SDSS\,J0100+1818 one of the few galaxy-scale systems known to date with three lensed sources at different redshifts \citep[see e.g.,][]{Collett2020}.

\section*{Acknowledgements}
This research is based on observations made with the NASA/ESA Hubble Space Telescope obtained from the Space Telescope Science Institute, which is operated by the Association of Universities for Research in Astronomy, Inc., under NASA contract NAS 5–26555. These observations are associated with program GO-15253. Based on observations collected at the European Organisation for Astronomical Research in the Southern Hemisphere under ESO programme 091.A-0852(A), and on observations with the Nordic Optical Telescope, operated by the Nordic Optical Telescope Scientific Association at the Observatorio del Roque de los Muchachos, La Palma, Spain, of the Instituto de Astrofisica de Canarias. The data presented were obtained with ALFOSC, which is provided by the Instituto de Astrofisica de Andalucia (IAA) under a joint agreement with the University of Copenhagen and NOTSA. This project is partially funded by PRIN-MIUR 2020SKSTHZ. RC and SHS thank the Max Planck Society for support through the Max Planck Research Group for SHS.

 
\bibliographystyle{aa} 
\bibliography{bibliografia} 

\begin{appendix}

\section{Spectra of the multiple images}
\label{app_a}

We show here the 1D spectra of the multiple images of the AB source that are covered by the X-Shooter data (as described at the beginning of Section~\ref{sec:system}). 
\begin{figure}[!h]
   \centering
   \includegraphics[width=7cm]{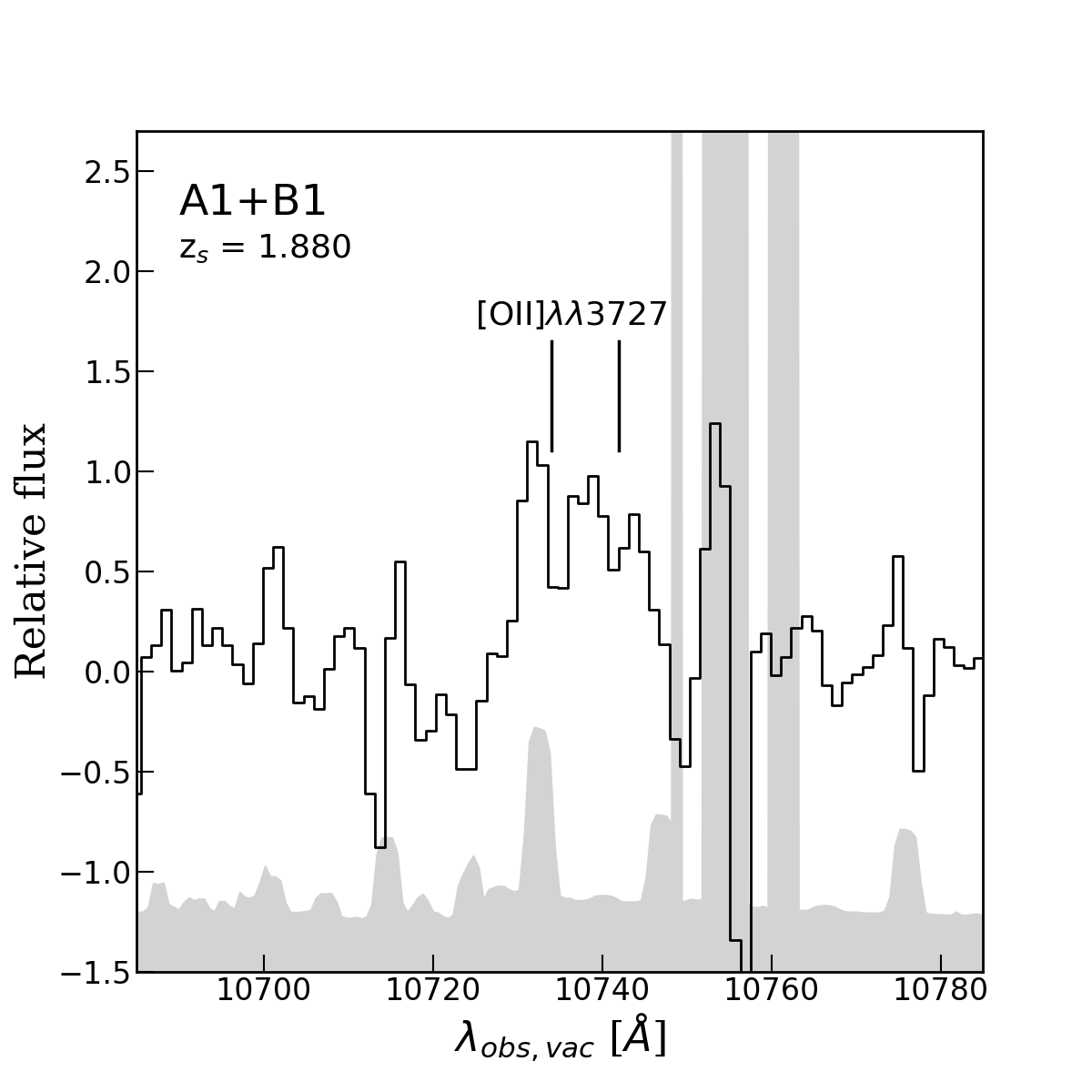}
      \caption{X-Shooter 1D spectra of the multiple images A1 and B1. Zoom in of the wavelength range where we observe the main emission lines used to confirm the redshift measurement. The black lines are the observed spectra in units of $10^{-17}$ erg s$^{-1}$cm$^{-2}$ \AA$^{-1}$, and the grey regions indicate the data variance.}
         \label{fig:spectrum_A1B1}
   \end{figure}
   
\begin{figure}[!h]
   \centering
   \includegraphics[width=7cm]{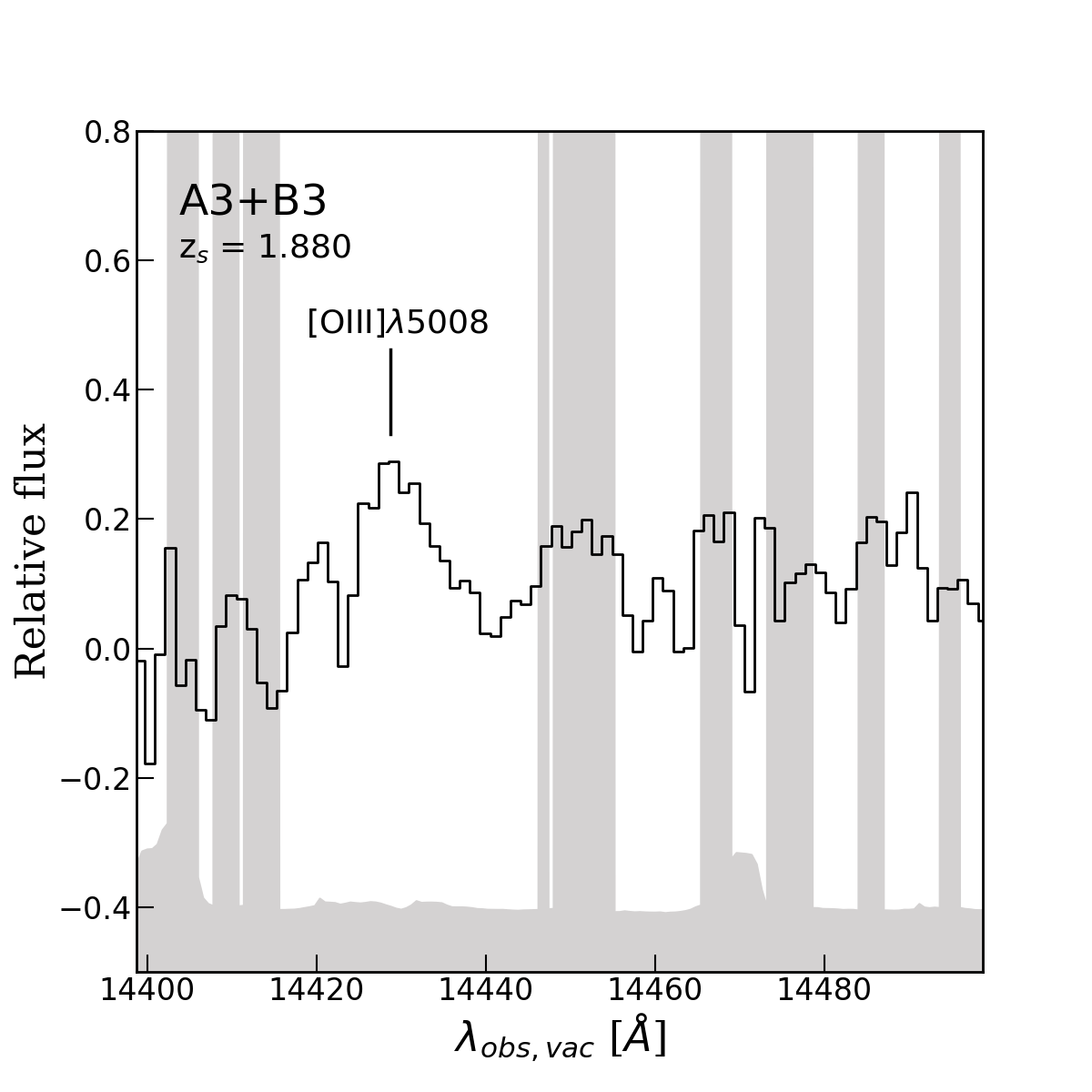}
      \caption{X-Shooter 1D spectra of the multiple images A3 and B3. See the caption of \ref{fig:spectrum_A1B1} for further details.}
   \end{figure}
   
\begin{figure}[!h]
   \centering
   \includegraphics[width=7cm]{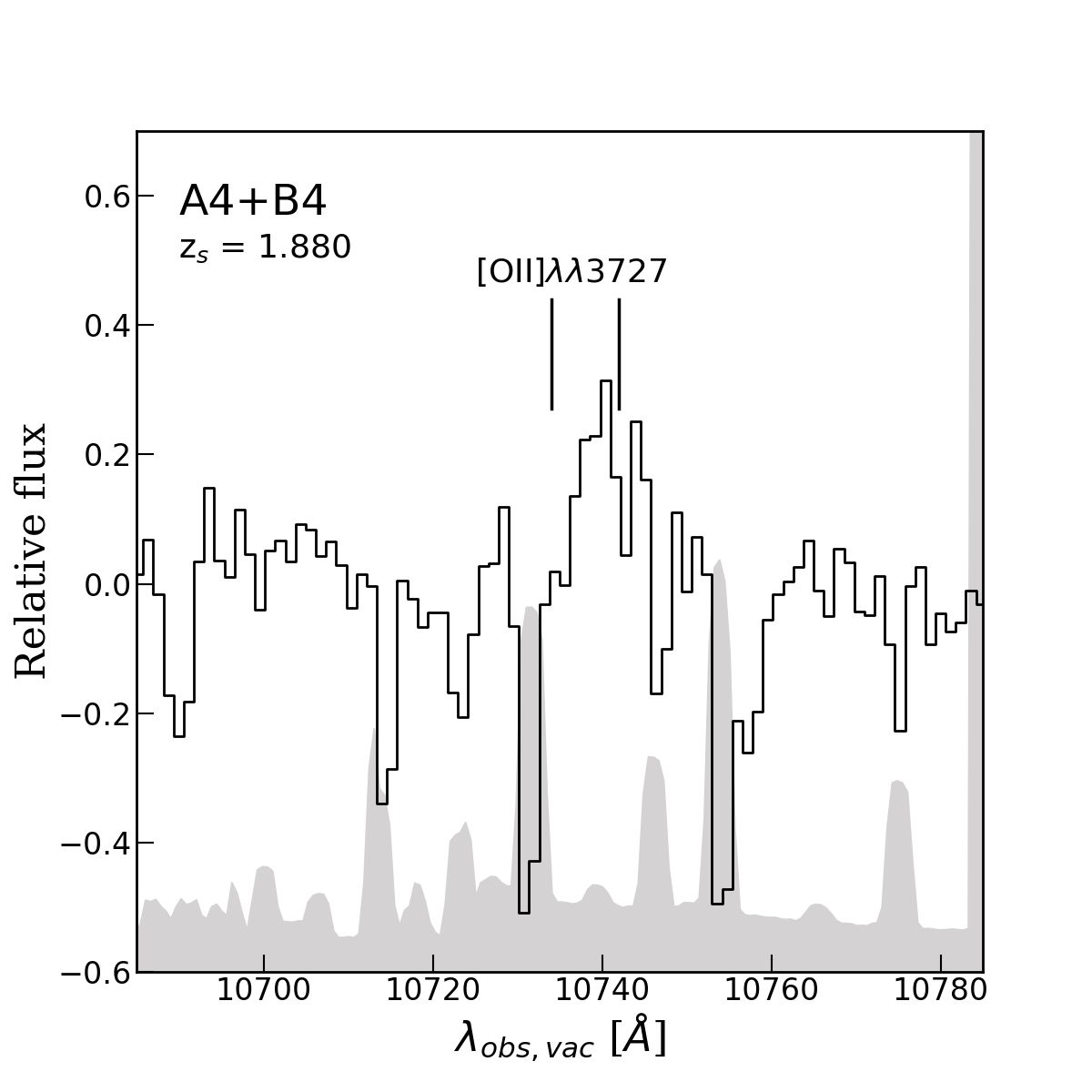}
      \caption{X-Shooter 1D spectra of the multiple images A4 and B4. See the caption of \ref{fig:spectrum_A1B1} for further details.}
   \end{figure}
   
\section{Spectrum of the main deflector}
\label{app_b}
We show here the 1D spectrum of the main lens galaxy. 
\begin{figure*}[!h]
   \centering
   \includegraphics[width=0.8\textwidth]{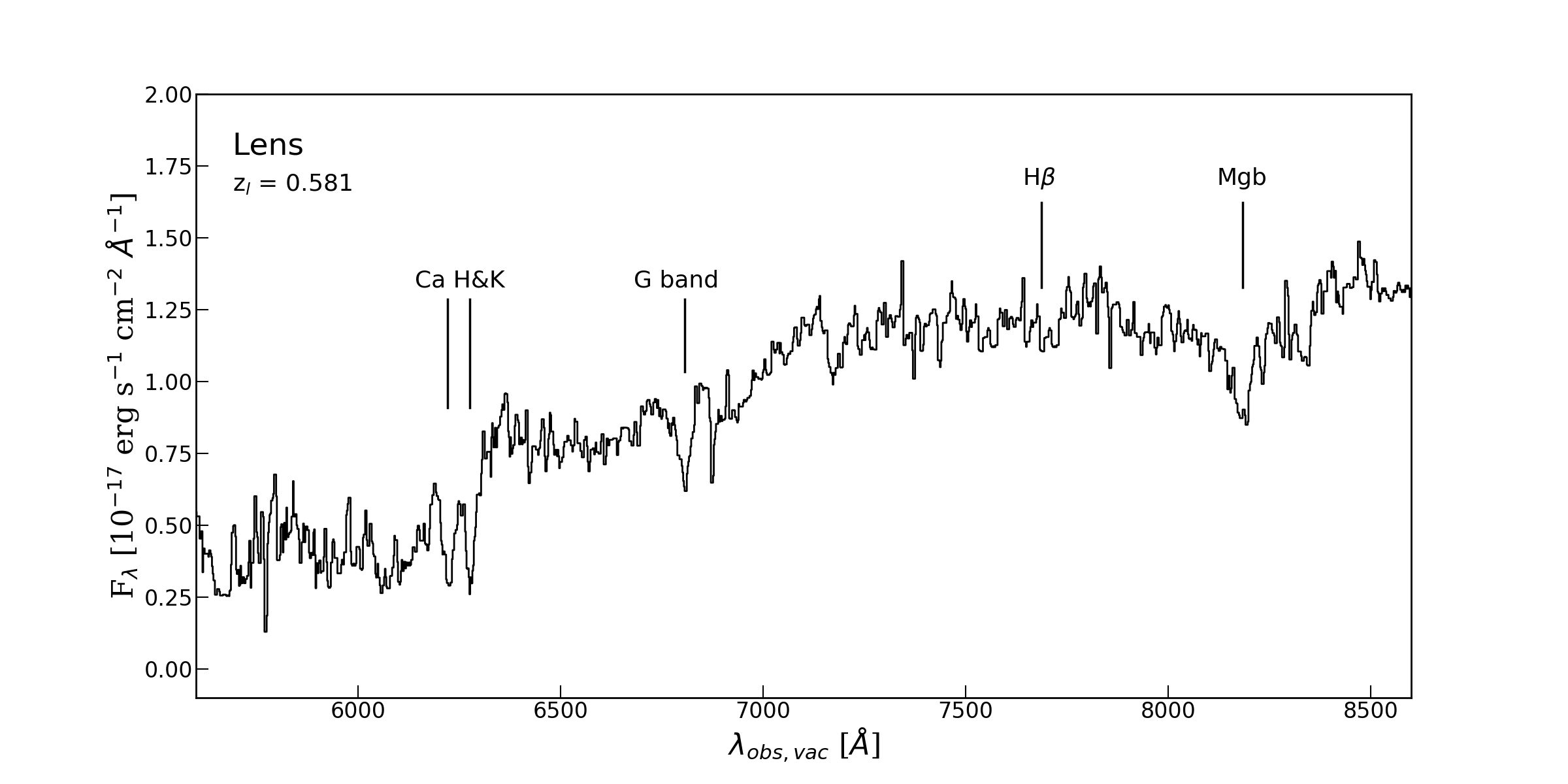}
   \caption{1D X-Shooter spectrum of the main lens galaxy. It is smoothed with a boxcar filter, and the most prominent spectral features are marked.}
              \label{fig:spectrum_lens}%
    \end{figure*}
\end{appendix}
\end{document}